\documentclass[rmp,aps,floatfix,twocolumn,square,unsortedaddress]{revtex4}
\usepackage{graphicx}

\begin{document}

\title{The Canonical Function Method and its applications in Quantum Physics}

\author{C. Tannous\dag,  K. Fakhreddine* \mbox{  }  and J. Langlois\S}
\affiliation{\dag Laboratoire de Magn\'etisme de Bretagne, CNRS-FRE 2697, \\
\S Laboratoire des Collisions Electroniques et Atomiques,\\
Universit\'e de Bretagne Occidentale, BP: 809 Brest CEDEX, 29285 France \\
*Faculty of Science, Lebanese University and CNRS, 
P.O. Box: 113-6546, Beirut, Lebanon}

\begin{abstract}
The Canonical Function Method (CFM) is a powerful method that solves
the radial Schr\"{o}dinger equation for the eigenvalues directly without having to evaluate the
eigenfunctions. It is applied to various quantum mechanical problems in Atomic
and Molecular physics with presence of regular or singular potentials. It has also been
developed to handle single and multiple channel scattering problems where the phaseshift
is required for the evaluation of the scattering cross-section.
Its controllable accuracy makes it a valuable tool for the evaluation of vibrational levels
of cold molecules, a sensitive test of Bohr correspondance principle and a powerful method
to tackle local and non-local spin dependent  problems.
{\em To submit to Rev.\ Mod.\ Phys.\ (2006).}
\end{abstract}

\maketitle

\setcounter{tocdepth}{4}
\tableofcontents

\section{Introduction}

The Canonical Function Method (CFM) is a powerful means for solving the Radial Schr\"{o}dinger
Equation (RSE), a singular boundary value problem (SBVP) with a stringent dual requirement
of regularity near the origin ($r \sim 0$) where the potential is large and 
near infinity ($r \rightarrow \infty$) where the potential is very small.

The CFM turns this SBVP into a regular initial value problem and allows full and accurate
determination of the spectrum of the Schr\"{o}dinger operator.

The CFM can handle bound states and scattering states, spinless and spin-dependent, 
single channel or multi-channel problems and possesses several features that makes it of 
interest for solving a large variety of
problems in Quantum Mechanics of Atoms, Molecules and Scattering. Those capabilities are based
essentially on the following main characteristics:

\begin{itemize}
\item Its capability to transform a SVBP into a regular initial value problem (RIVP),
means mathematically that the CFM belongs to a family of Invariant Embedding Methods (IEM).
Those essentially transform a boundary value problem into an initial value problem. 
The CFM is even more powerful than standard IEM techniques since it turns the nature of 
the problem from singular to regular.
\item The evaluation of the Schr\"{o}dinger operator spectrum is done without performing
diagonalization, bypassing the evaluation of the eigenfunctions. This
allows to preserve a high degree of numerical precision that is required in solving sensitive
eigenvalue problems arising in Cold Molecules or large $n$-limit problems where levels to evaluate are
very close to the continuum limit or dissociation. 
\end{itemize}

The numerical precision gained with the bypass of intermediate digonalization operations
is reminiscent of the Golub-Reinsch algorithm (see for instance ref.~\cite{Recipes})
used for the singular value decomposition of arbitrary rectangular matrices.

This review is organised as follows: The next section is a description of the
CFM with highlights of its mathematical aspects and its generalisation to
the multichannel case. In section III the mathematical
and numerical techniques of the radial Schr\"{o}dinger equation are discussed
and the full procedure for dealing with general potentials, energy levels
and evaluation of phaseshifts is detailed in section IV.
In the rest of the sections, we apply the CFM method systematically to a variety of
problems highlighting its successes and showing its limitations whenever they
show up. While, the versatility of the method is illustrated with the help of 
this selection of examples, the list below is by no means exhaustive but serves the purpose
of underlining the breadth of applications of the CFM. 

\begin{itemize}
\item Energy levels for regular and singular potentials (radial case)
\item Potential estimation (in parametric or pseudo form) from spectroscopic data
\item Test of the Bohr Correspondence Principle
\item Vibrational energy levels of Cold Molecules and application to the $^{23}{\rm Na}_2$
molecule in the $0^-_g$ and $1_u$ electronic states. The Lennard-Jones molecule case is also
studied.
\item Local and  Non-Local Exchange problems.
\item Accurate Phase shift evaluation for regular and singular potentials.
\end{itemize}

In the last section before the conclusion, we treat the general but non obvious 
(regular and singular) 1D potentials to illustrate how the method can be applied in this 
case of lesser interest than the 3D's.

In the Appendix we provide information linking the various units used since the review
contains results spanning several fields of physics where different systems of units are used.
The boundary conditions affecting the spectrum determination via the matching conditions
at a single radial point are also discussed in the Appendix.

\section{The Canonical Function Method (CFM)}

The CFM \cite{Kobeissi82} is a powerful means for solving the Radial Schr\"{o}dinger
Equation (RSE). The mathematical difficulty of the RSE lies in the fact it is a singular boundary value 
problem. The CFM turns it into a regular initial value problem and allows the full determination of the 
spectrum of the Schr\"{o}dinger operator bypassing the evaluation of the eigenfunctions.\\

The partial wave form of the RSE is written as:

\begin{equation}
-\frac{\hbar^2}{2\mu}{\frac{d^2 u_{l} ( E; r )}{d r^2}}  + 
\left[{V( r ) +  \frac{\hbar^2}{2\mu}\frac{ l (l + 1 )}{r^2}}\right] u_{l} (E; r ) = E u_{l} (E; r )
\end{equation}

where $\mu$ is the reduced mass and $u_{l} (E; r )$ is the reduced probability amplitude for 
orbital angular momentum $l$ and eigenvalue $E$.

The boundary conditions are:
\begin{equation}
\lim_{r \rightarrow 0} u_{l} (r) =0; \lim_{r \rightarrow +\infty} u_{l} (r) =0
\end{equation}

The CFM developed initially by Kobeissi \cite{Kobeissi91} and his coworkers to integrate the RSE, 
consists of writing the general  solution $y(r)$ representing the probability
amplitude $u_{l} (E; r )$ as a function of the radial distance 
$r$ in terms of two basis functions $\alpha(E;r)$ and $\beta(E;r)$ for some energy $E$. \\
Generally, the RSE is rewritten in a system of units such that $\hbar=1, 2 \mu=1$ 
(see Appendix on units):
\begin{equation}
\frac{d^2 y(r)}{d r^2} = \left[{V( r ) +  \frac{ l (l + 1 )}{r^2}-E }\right] y(r) 
\label{schro}
\end{equation}
At a selected distance $r_0$, a well defined set of initial conditions are satisfied by the 
canonical functions and their derivatives ie: $\alpha(E;r_0)=1$ with $\alpha'(E;r_0)=0$ 
and $\beta(E;r_0)=0$ with $\beta'(E;r_0)=1$. Thus we write:

\begin{equation}
y(r)= y(r_0) \alpha(E;r)  + y'(r_0)\beta(E;r)
\end{equation}

The method of solving the RSE is to proceed from $r_0$ simultaneously towards the origin
($r \rightarrow 0$) and towards infinity ($r \rightarrow \infty$). 
Expressing the continuity condition of the "wavefunction" $y(r)$ and its derivative
$y'(r)$ at the point $r_0$ from the left ($r \rightarrow 0$) and right ($r \rightarrow \infty$)
yields (see Appendix on matching and boundary conditions):

\begin{eqnarray}
\left.\frac{y'(r_0)}{y(r_0)}\right]_{-} =  -\frac{\alpha(E;0)}{\beta(E;0)} \nonumber \\
\left. \frac{y'(r_0)}{y(r_0)}\right]_{+} =  -\frac{\alpha(E;\infty)}{\beta(E;\infty)}
\end{eqnarray}

\begin{figure}[htbp]
\begin{center}
\scalebox{0.7}{\includegraphics{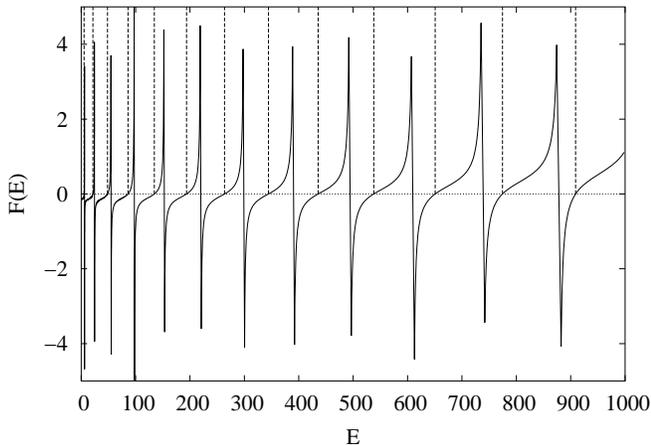}}
\end{center}
\caption{Typical behavior of the eigenvalue function with energy. The vertical lines indicate
 the eigenvalue position. The graph of the eigenvalue function has an approximate
 $\tan(E)$ shape versus the energy $E$.}
\label{tan}
\end{figure}

When the integration is performed, the ratio of the $r$ dependent canonical functions is monitored until
saturation with respect to $r$ is reached at both limits ($r \rightarrow 0$ and $r \rightarrow \infty$).
The saturation of the $\frac{\alpha(E;r)}{\beta(E;r)}$ ratio with $r$ yields a position independent 
eigenvalue function $F(E)$. \\
The latter is mathematically defined with the help of two associated energy functions:
\begin{equation}
l_{-}(E)= \lim_{r \rightarrow 0} -\frac{\alpha(E;r)}{\beta(E;r)} 
\end{equation}
and:
\begin{equation}
l_{+}(E)= \lim_{r \rightarrow +\infty} -\frac{\alpha(E;r)}{\beta(E;r)} 
\end{equation}

as:

\begin{equation}
F(E)=l_{+}(E)-l_{-}(E)= \left[ \frac{y'(r_0)}{y(r_0)}\right]_{+} - \left[\frac{y'(r_0)}{y(r_0)}\right]_{-}
\end{equation}

its zeroes expressing the continuity of $y(r)$ and its derivative $y'(r)$ at the point
$r_0$ (matching conditions) yield the spectrum of the RSE. In practice we have:

\begin{equation}
l_{-}(E) \approx  -\frac{\alpha(E;r_{min})}{\beta(E;r_{min})}
\end{equation}
and:
\begin{equation}
l_{+}(E) \approx -\frac{\alpha(E;r_{max})}{\beta(E;r_{max})} 
\end{equation}

with $r_{min}, r_{max}$  the radial coordinates where saturation is observed (within
some predefined tolerance) respectively in  $l_{-}(E)$ and $l_{+}(E)$.
An example of typical behaviour of $F(E)$ is
displayed in Fig. ~\ref{tan}. The eigenfunctions may be obtained for any $E=E_k$ 
where $E_k$ is a zero of $F(E)$.\\

The eigenvalue function definition depends on the type of boundary conditions at hand. 
In the Appendix we describe the general boundary conditions case and the corresponding
matching conditions along with the corresponding eigenvalue function. \\

The $\tan(E)$ shape of $F(E)$ provides a deep insight into the physical significance
of the CFM method. The latter transforms a SBVP from the open interval $[0, \infty[$ 
to the finite interval $[r_{min}, r_{max}]$ defined by the saturation coordinates of the
energy functions. This means the CFM maps an arbitrary potential $V(r)$ onto the
infinite square well problem defined by $V(r)=0$ in the open interval $]r_{min}, r_{max}[$
and $V(r_{min})= V(r_{max})=\infty$ (see section VII for additional information). In fact, for
the infinite square well problem $F(E)$ has a $\tan(E)$ (see also ref.~\cite{Johnson}). 

The speed and accuracy of the CFM method have been tested and compared to standard
integration algorithms  such as the order four Runge-Kutta (RK4) method, 
Numerov etc.. in a variety of cases and for a wide of range of potentials. 

More specifically,  $r_0$ being selected as the starting point for the
integration, the $r$-axis is divided into intervals $I_p =[r_p, r_{p + 1}]$
in a way such that the potential can be series expanded over $I_p$ 
(see refs~\cite{Fornberg1,Fornberg2}). \\
The expansion uses coefficients $\gamma_{n}^{(p)}$  so that the local
expression of the total potential is defined as: 
\begin{equation}
\left[V(r)+  \frac{ l (l + 1 )}{r^2}\right]_{r \in I_p}=  \sum_{n = 0}^{\infty} \gamma_{n}^{(p)} x^n, x \in I_p  
\end{equation}

In order to perform integration, we use the following variable step difference equation: 
 
\begin{equation}
  y_{p + 1} = y_{p} + h_{p} {y'}_p + \sum_{n = 2}^{\infty} C_n^{(p)} {h}_p^n
\end{equation}

\begin{equation}
  {y'}_{p + 1} = {y'}_p+ \sum_{n = 2}^{\infty} n {C}_n^{(p)} {h}_p^{n - 1}
\end{equation}

where the ${y}_p = {y}(r)$ are a particular set of canonical
functions and the ${y'}_p = {y'}(r)$ are their
derivatives in the interval $[r_p, r_{p+1}]$. 
The $C_n^{(p)}$ are given by the recursion formula for energy $E$ obtained from 
series expanding eq.~\ref{schro}:
\begin{equation}
 (n + 2) (n + 1) C_{n + 2}^{(p)} = \sum_{m = 0}^{n} C_m^{(p)} \gamma_{n - m}^{(p)} - E C_n^{(p)}
\end{equation}
with $C_0^{(p)} = y_p$ and $C_1^{(p)} = {y'}_p$.

In practice, the above sums over $n$ are truncated to a cutoff value $N$ (instead of $\infty$)
chosen large enough so that the remainder error is less than a selected 
truncation error $\epsilon_T$ . \\
 Given $N$ and $\epsilon_T$, the integration step $h_p$ for the interval 
 $[r_p, r_{p+1}]$ is deduced from 
$ h_p = {(\epsilon_T^2 / |{C}_n^{(p)}|)}^{\frac{1}{N}} $. \\
At the starting point of integration $r_0$, the starting step size $h_0$ is thus determined, 
and the procedure is repeated at $r_1=r_0 +h_0$ leading to $h_1$ and so forth 
until we reach  $r_{p+1}=r_p +h_p$ after $p$ steps. \\

These equations allow the propagation of the solution from one point to the
next using a variable step and a local series expansion 
of variable order controlled by the truncation error $\epsilon_T$. 
The accuracy of the results are monitored
with respect to the decrease of $\epsilon_S$ and  the procedure is referred to as the  
VSCA (Variable Step with Controlled Accuracy) method.

Generally, one avoids using or evaluating the radial wavefunction but if one insists on
evaluating it, the canonical functions $\alpha(E;r)$ and $\beta(E;r)$ are used to
determine the radial wavefunction at any energy via the expression:

\begin{equation}
u_l(E;r)=u_l(E;r_0) \alpha(E;r) + u'_l(E;r_0) \beta(E;r)
\end{equation}
where  $u_l(E;r)$ and $u'_l(E;r_0)$ are the radial wavefunction and its derivative at
the initial distance $r_0$. 

Before proceeding any further, the essential test of any method that pretends solving
the RSE is the Coulomb potential and the essential case to test the accuracy and reliability 
of the CFM is the (textbook) Hydrogen atom.

The CFM results are shown in Table.~\ref{coul} along with the exact analytical
results.

\begin{table}[!ht]
\begin{center}
\begin{tabular}{ c c c }
\hline
Index& CFM (Ry) & Exact (Ry) \\
\hline
 1   &-1.00000   &   -1.00000 \\
   2  &-0.250000    &  -0.250000 \\
   3  &-0.111111    & -0.111111 \\
   4 &  -6.25000(-2)    &  -6.25000(-2) \\
   5  & -4.00000(-2)    &   -4.00000(-2) \\
   6  & -2.77778(-2)    &   -2.77778(-2) \\
   7  & -2.04082(-2)    &   -2.04082(-2) \\
   8 &  -1.56250(-2)    &   -1.56250(-2) \\
   9  & -1.23457(-2)    &   -1.23457(-2) \\
   10 &  -1.00000(-2)    &   -1.00000(-2) \\
   11 &  -8.26446(-3)    &   -8.26446(-3) \\
   12  & -6.94444(-3)    &   -6.94444(-3) \\
   13  & -5.91716(-3)    &   -5.91716(-3) \\
   14 &  -5.10204(-3)    &   -5.10204(-3) \\
   15 &  -4.44445(-3)    &   -4.44445(-3) \\
   16 &  -3.90625(-3)    &   -3.90625(-3) \\
   17 &  -3.46021(-3)    &   -3.46021(-3) \\
   18 &  -3.08642(-3)    &   -3.08642(-3) \\
   19 &  -2.77008(-3)    &   -2.77008(-3) \\
   20 &  -2.50000(-3)    &   -2.50000(-3) \\
   21 &  -2.26757(-3)    &   -2.26757(-3) \\
   22 &  -2.06612(-3)    &   -2.06612(-3) \\
   23 &  -1.89036(-3)    &   -1.89036(-3) \\
   24 &  -1.73611(-3)    &   -1.73611(-3) \\
\hline
\end{tabular}
\end{center}
\caption{Energy levels of the Hydrogen atom. Middle column values are the CFM results whereas
the last column values are the corresponding exact analytically obtained values. 
The numbers in parenthesis represent a power of 10.}
\label{coul}
\end{table}

It is remarkable to notice that all digits (calculated by CFM and analytically) are
all same.\\

The classical Morse potential is the simplest model for the evaluation of vibrational spectra
of  diatomic molecules. First of all,  we have
analytical expressions for the levels, secondly it provides a stringent test of the CFM before embarking
into more sophisticated cases such as weakly bound or cold molecules (also called long-range molecules).

The Morse potential is given by:

\begin{equation}
        V(r)=D {[ 1  -\exp(-a \{r-r_e \}) ]}^2 -D
\end{equation}

with the values $D,a,r_e$ equal respectively to  188.4355,  0.711248,   1.9975 in au.
The analytic expression for the levels is:

\begin{equation}
E_{n}=- \frac{a^2 \hbar^2}{2 \mu}(\sqrt{2\mu D}/a -n-1/2)^2,
\end{equation}

with max $n \le \sqrt{2\mu D}/a -1/2$. Hence the number of levels
 is given by: $N=\sqrt{2\mu D}/a-1/2$.

Working with units such that $\hbar=1$ and $2\mu=1$,
the Morse potential and the eigenvalue function $F(E)$ are displayed in the figures below.
The table contains the levels calculated by CFM and compared to the analytical case.
Again, like in the pure Coulomb case treated previously, the agreement is perfect
and we found all the levels ($N=19$) as predicted analytically.

\begin{figure}[htbp]
\begin{center}
\scalebox{0.5}{\includegraphics{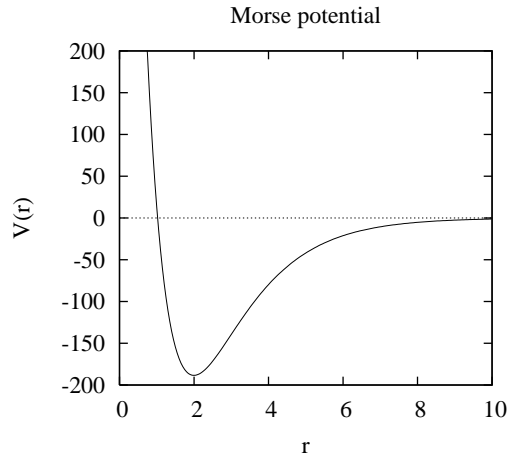}}
\end{center}
\caption{Morse potential  $V(r)=D {[ 1  -\exp(-a \{r-r_e \}) ]}^2 -D$ with parameters $D= 188.4355, a= 0.711248,  r_e= 1.9975$.} 
\label{morsepot}
\end{figure}

\begin{table}[!ht]
\begin{center}
\begin{tabular}{c c c}
\hline
Index & CFM  & Exact \\
\hline
 1 & -178.798248 &  -178.798538 \\
 2 & -160.282181 &  -160.283432 \\
 3 & -142.778412 &  -142.78006 \\
 4 &  -126.287987 &  -126.288445 \\
 5 &  -110.807388 &  -110.808578 \\
 6 &  -96.3395233  & -96.3404541 \\
 7 &  -82.8832169 &  -82.884079 \\
 8  & -70.4389801 &  -70.4394531 \\
 9 & -59.0056    &  -59.0065727 \\
 10 & -48.5851288 &  -48.5854378 \\
 11 &  -39.1754532 &  -39.1760521 \\
 12 & -30.77771  &   -30.7784157 \\
 13 & -23.3919983 &   -23.3925247 \\
 14 & -17.0183048  &  -17.018383 \\
 15 & -11.6557436 &   -11.6559868 \\
 16 &  -7.3050122 &  -7.30533791 \\
 17 & -3.9661877 &   -3.9664371 \\
 18 & -1.6390723  &   -1.63928342 \\
 19 & -0.3238727 &    -0.32387724 \\
 \hline
\end{tabular}
\end{center}
\caption{Energy levels of the Morse potential $V(r)=D {[ 1  -\exp(-a \{r-r_e \}) ]}^2 -D$ with parameters $D= 188.4355, a= 0.711248,  r_e= 1.9975$. Middle column values are the CFM 
results whereas the last column values are the corresponding exact analytically obtained values.
Units are such that $\hbar=1$ and $2\mu=1$.}
\label{morset}
\end{table}

\begin{figure}[htbp]
\begin{center}
\scalebox{0.5}{\includegraphics*{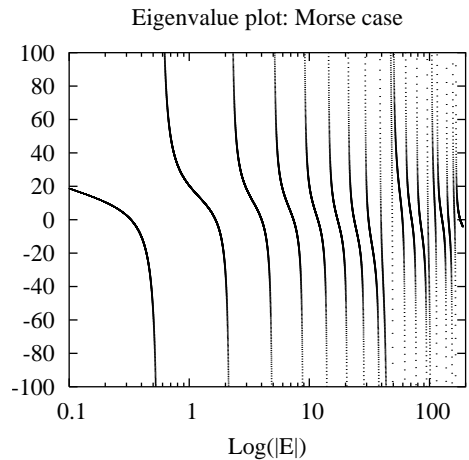}}
\end{center}
\caption{Behavior of the eigenvalue function $F(E)$ with energy on a
semi-log scale for the Morse potential.} 
\label{morse}
\end{figure}

\subsection{The Multichannel case}

The CFM can be extended to the multichannel (or multi-component) problem, where the functions
$\alpha(E;r)$ and $\beta(E;r)$ are no longer scalar functions of $E$ and $r$. 
Mulichannel case  occurs when we have a potential well with several local minima or when
we are dealing with Scattering problems requiring simultaneously incident  and
scattered wavefunctions. In addition, spin leads naturally to multicomponent wavefunctions. \\

Starting from a Schr\"odinger system of equations of the form:

\begin{equation}
\sum_{k=1}^{N}F_{ik}y_{k}(r)=0, \mbox{         } i=1,2,...,N
\end{equation}

where:

\begin{eqnarray}
F_{ik}(r) & = & (2 \mu/\hbar^2)V_{ik}(r)    \nonumber  \\
F_{ii}(r) & = & (2 \mu/\hbar^2)(-d^2/dr^2 -E_{i}+V_{ii}(r))
\end{eqnarray}

The elements $F_{ik}(r)$ of the matrix $F$ are physical operators and the
unknown functions $y_{k}(r)$ are partial radial waves satisfying the 
boundary conditions at $r=0$ and $r=\infty$:

\begin{equation}
y_{k}(r=0)=y_{k}(r=\infty)=0 \mbox{      }\forall k
\end{equation}

Following Friedman and Jamieson \cite{Friedman}, we rewrite the $N$-component equation 
as a set of $N$ coupled Schr\"{o}dinger equations for a matrix potential with elements
$V_{ij}(r)$: 
\begin{equation}
y_{i}"(r) + (2 \mu/\hbar^2)[ E- V_{ii}(r)] y_{i}(r)= (2 \mu/\hbar^2) \sum_{k=1}^{N}V_{ik}y_{k}(r)
\end{equation}

where $i,k=1,2,...,N$. \\

Using linear superposition, we can write the full solution as:

\begin{equation}
y(r)= \sum_{k=1}^{N}[y_{k}(r_0)\alpha_{ik}(E; r) + y'_{k}(r_0)\beta_{ik}(E; r)]
\end{equation}

where, as before, $r_0$ is an arbitrary point between zero and infinity,
$\alpha_{ik}(E; r), \beta_{ik}(E; r)$ are $2N$ independant particular solutions
satisfying the set of initial conditions:

\begin{eqnarray}
\alpha_{ik}(E; r_0) = &  \beta'_{ik}(E; r_0) = & \delta_{ik}  \nonumber  \\
\alpha'_{ik}(E; r_0) = &  \beta_{ik}(E; r_0) = & 0
\end{eqnarray}

Going from the indicial to the matrix representation, we can build a matrix 
$\bf{L}(E,r)$ that depends on the energy $E$ and radial position $r$ defined by:

\begin{equation}
\bf{L}(E,r)=\bf{\beta}^{-1}(E,r)\bf{\alpha}(E,r)
\end{equation}

where $\bf{\alpha}$ and $\bf{\beta}$ are $N \times N$ matrices whose elements are
the $\alpha_{ik}(E; r), \beta_{ik}(E; r)$ functions.

The analogue of the eigenvalue function is now given by the determinant equation:
\begin{equation}
D(E)=| \bf{L}(E,0) - \bf{L}(E,\infty)|
\end{equation}

The eigenvalues are given by the zeroes of the determinant $D(E)=0$. \\

Let us apply the above to the particular case where the matrix potential $\boldmath{V(r)}$ 
\cite{Friedman} is a constant coupling matrix $\bf{C}$ multiplying a scalar function $M(r)$:

\begin{equation}
\boldmath{V(r)}=\boldmath{CM(r)}, \hspace{2mm} V_{ij}=C_{ij}M(r), \hspace{2mm} i,j=1,..6
\label{coupled}
\end{equation}

The $r$ dependence is contained solely in the radial function $M(r)=z^2-2z, z=\exp(-ax), x=r-r_e$ 
where $r_e$ is the scalar potential (minimum) equilibrium value.
$\boldmath{C}$ is a full (6$\times$6) constant matrix that may be diagonalised by the
similarity transformation: $\boldmath{C}=\boldmath{Q}\boldmath{G}\boldmath{Q}^{-1}$. \\
$\boldmath{Q}$ is a non-singular real constant matrix and $\boldmath{G}$ is a (6$\times$6) diagonal
matrix. Picking a coupling matrix $C$ similar to a diagonal matrix $G$  whose elements 
$\boldmath{G}_{ii}, i$ =1..6  are respectively: $\{24,31.5,37.5,42,56,63\}$, the initial
fully coupled six-channel problem~\ref{coupled} is transformed into six uncoupled 
Schr\"{o}dinger equations that are solved with the multichannel CFM. The full spectrum
of $\boldmath{V(r)}$ is made from the union of the sets of eigenvalues of the six individual
Schr\"{o}dinger equations. \\

In order to benchmark the CFM results, we note that this multichannel example 
is nothing other than a special matrix form of the Morse potential. Then,
we have access to the full spectrum analytically as in the scalar Morse case: 

\begin{equation}
E_{in}=-w_e x_e (d_i-n-1/2)^2, \mbox{   } i=1,...6,  n \le d_i-1/2,
\end{equation}

where $d_i=\sqrt{(G_{ii}/w_e x_e)}$ and $w_e x_e=a^2 \hbar^2 /2 \mu$.
In order to perform a detailed numerical comparison between the 
numerical performance of the CFM
and the multichannel Morse potential analytical results, we take: 
$r_e=1.5\AA, w_e x_e=8$ cm$^{-1}$ and $a=1.540$ \AA. \\
The comparison between the computed  and the exact eigenvalues displayed
in  table~\ref{multi} shows that they are almost indistinguishable.

\begin{table}[!ht]
\begin{center}
\begin{tabular}{c  c  c}
\hline
$(i,n)$ & Exact  & CFM\\
\hline
(6,0) & -42.550 055 679 356  & -42.550 055 679 344  \\
(5,0) & -36.833 989 511 483   & -36.833 989 511 484  \\
(4,0) & -25.669 697 220 176  & -25.669 697 220 173  \\
(3,0) & -22.179 491 924 311   & -22.179 491 924 313  \\
(2,0) & -17.625 492 133 612   & -17.625 492 133 614  \\
(6,1) & -13.650 167 038 069   & -13.650 167 038 068  \\
(1,0) & -12.143 593 539 448   & -12.143 593 539 445 \\
(5,1) & -10.501 968 534 449   & -10.501 968 534 446 \\
(4,1) & -5.009 091 660 529  &    -5.009 091 660 528 \\
(3,1) & -3.538 475 772 933   &   -3.538 475 772 933  \\
(2,1) & -1.876 476 400 837   &   -1.876 476 400 834  \\
(6,2) & -.750 278 396 781   &    -.750 278 396 781  \\
(1,1) & -.430 780 618 346   &    -.430 780 618 346  \\
(5,2) & -.169 947 557 416   &   -.169 947 557 416  \\ 
\hline     
\end{tabular}
\end{center}
\caption{Analytical and Multi-channel CFM computed eigenvalues $E_{in}$.
All values are in cm$^{-1}$.}
\label{multi}
\end{table}

\subsection{Evaluation of energy spectra and phaseshift}

Before treating any problem with the CFM a number of constraints should be underlined
in order to properly tackle any problem with this method whether it pertains to
energy spectra evaluation or phaseshift:

\begin{itemize}

\item General considerations:

\begin{itemize}

\item The potential is spherically symmetric.

\item  The interval of interest should be of infinite length.

\item  Ill-conditioning and accuracy:
The spectrum depends on the zeroes of $F(E)=l_{+}(E)-l_{-}(E)$.
This subtraction leads to inaccuracies because the entire spectrum
depends on the zeroes of $F(E)$ in the single channel case 
or $D(E)=| \bf{L}(E,0) - \bf{L}(E,\infty)|$ in the multichannel case.

\item $r_{0}$ issue and the number of eigenvalues:
The number of eigenvalues depend strongly on $r_{0}$.
$r_{0}$ should be increased until a $\tan(x)$ like diagram for the energy
function is obtained. \\
For short-range potentials, $r_{0}$ should be close to the minimum of the potential 
or the equilibrium value of the radial distance $r_{e}$.
If $r_{0}$ is far from minimum, it will see only those levels close
to the value of $V(r_{e})$.

\end{itemize}

\item Energy spectra calculations:

\begin{itemize}

\item The method being  sensitive to convergence for $r\rightarrow 0$ and $r\rightarrow \infty$,
One has to check that for $r \rightarrow 0$ the canonic functions $\alpha$
and $\beta$ are diverging near the origin in the same way (saturation of 
the ratio).

\item  One has to check that for $r \rightarrow \infty$ the canonic functions
 $\alpha$ and $\beta$ have reached their Asymptotic behaviour (of the Coulombic
form  $\sin(kr)/kr$).

{\bf Note:} This is not true for the phase.

\item One should have a regular structure of the $\tan(E)$ type for the 
energy function $F(E)=l_{+}(E)-l_{-}(E)$ in the single channel 
(or $D(E)=| \bf{L}(E,0) - \bf{L}(E,\infty)|$ in the multichannel) case. 

\item The method is very sensitive to $r_{0}$.

\item Forward integration step $r \rightarrow \infty$: typically it is 0.1 to 1 in energy
calculations. In phase calculations it is 0.1 to 0.01. It is sensitive
to the energy used. When the energy increases the step should be decreased.
There is a compromise to be reached because if the step is too much reduced
any physical quantity will saturate immediately.

\end{itemize}

\item Phaseshift calculations:

\begin{itemize}

\item The phase stability might be reached long before the asymptotic
regime in $\Psi(E; r) \sim \sin(kr)/kr$ is reached.
Typically the phase is stable when the potential is about 0.1 au.

\item One has to check that for $r \rightarrow 0$ the canonic functions $\alpha$
and $\beta$ are diverging near the origin in the same way (saturation of 
the ratio).

\item The method being very sensitive to $r_{0}$, we ought to treat separately
the different types of potential. The general rules are enunciated below: 

\begin{itemize}
\item Short-range potentials:
Typically $r_{0} \sim 1 $ for all angular momentum $L$ and energies. 
For large energies $r_{0}$ should be reduced.
\item Long-range potentials:
Typically $r_{0} \sim ~ 1$ for all $L$ but for small energies (1-10 au). 
For large energies  $r_{0}$ should be reduced to 0.1.
\end{itemize}

\item Bessel functions are used for Short-range potentials only.
Trigonometric functions are used for Long-range and Short-range potentials potentials.

\item Forward integration step $r \rightarrow \infty$: typically it is 0.1 to 1 in energy
calculations. In phase calculations it is 0.1 to 0.01. It is sensitive
to the energy used. When the energy increases the step should be decreased.
There is a compromise to be reached because if the step is too much reduced
any physical quantity will saturate immediately.

\item The stability of the CFM is based on the following: two independent sets of
solutions are generated at some central point, integrating
inwards to the origin and outwards to the asymptotic region. Both contain
linear combinations of the regular and the irregular solutions, and by
suitably combining them, the irregular solution is eliminated. Integration
can be made out to a very large radius, allowing the phaseshift to be
determined by matching to plane or Coulomb wave solutions. It is not
necessary to obtain a series expansion of the solution in order to start the
integration, making it unnecessary to series expand the potential. \\

\item The method can be successfully applied to scattering problems where the
phaseshift is of paramount importance in the determination of the scattering 
cross section. Of particular importance are (e,2e) scattering problems that are 
dealt with in section VII.

\end{itemize}

\end{itemize}

A final point to consider when dealing with the CFM using standard integration methods
is that one cannot in general proceed toward the origin $r=0$
because of the potential singularity (for instance, the RK method blows up at $r=0$)
in contrast with the VSCA algorithm.

\section{Potential estimation from spectroscopy}
The high accuracy of the CFM allows to retrieve the potential (in parametric or pseudo form) from
the observed atomic or molecular spectra. We exploit the Quantum Defect (QD) as an
indicator of the results obtained by the CFM in order to find the optimal
parameters of the potential. \\
QD information is widely exploited in modern spectroscopy, to characterize
 Rydberg states and in the calculation of the photoionization cross sections of various 
 atomic and molecular species \cite{Jungen,Aymar96}. \\
We show in this section that QD information, used within the Distorted Wave Born 
Approximation (DWBA) framework~\cite{Whelan}, might also prove useful for the description of 
ionization processes by presenting an alternative way to account for the short 
range interactions (static and exchange) in the calculation of the final state 
continuum distorted waves. The range of validity of this approach reaches beyond 
that \cite{Riley} of the commonly used Furness-McCarthy local exchange 
approximation \cite{Furness}. Compared to the determination of the Hartree-Fock 
non-local operator \cite{Winkler}, which becomes rapidly a tedious task as the size
 of the target increases, our method allows for a target-independent procedure which
  can be readily applied to much larger atomic or molecular systems. \\

The parameters of the Green-Sellin-Zachor \cite{Green69} parametric form of the 
electron-ion potential are optimized in order to reproduce the QD using the 
CFM~\cite{Kobeissi82}. Several parametric potentials are discussed extensively 
in the literature and we already used several functional forms \cite{Aymar96} 
adapted to different atomic systems. We have studied \cite{Tannous99} the 
QD of some rare gases with the Klapisch parametric potential \cite{Klapisch71} 
and found that in some cases it was very difficult to optimize parameters that 
provide an accurate representation of the experimental QD. We believe, the 
Green-Sellin-Zachor \cite{Green69} is better suited to our present study as the parameter
space is small (two-dimensional) which allows for an efficient search of the optimized parameters. \\

The electron-ion potentials obtained for each Rydberg series are further modified to account
 classically for the electron-electron interaction in the final state. Calculations performed 
 within the DWBA framework for the ionization of argon in the equal energy sharing geometry
${\bf k}_a=-{\bf k}_b$, which is reasonably well documented both theoretically and 
experimentally, clearly validate our approach and shows significant improvements over previous treatments.

\subsection{Optimisation procedure}

An extensive review of the applications of model potentials has been given by Hibbert  \cite{Hibbert82}
and Aymar {\sl et al.} \cite{Aymar96}. The functional form suggested by Green {\sl et al.} \cite{Green69} is given by:
\begin{eqnarray}
&&V(r)= -(2/r) [(Z-1) \omega(r)+1],\nonumber\\
\mbox{with}\hspace{0.5cm}&& \omega(r)=1/[\epsilon_1{{\rm (exp}(r/\epsilon_2)-1})+1]
\label{eq5}
\end{eqnarray}
where $\epsilon_1$ and $\epsilon_2$ are parameters that are determined by the optimisation procedure.\\

The optimisation problem, at hand, is over-determined since the experimental set of energy levels 
might consist of tens of values whereas the potential depends only on two
numbers, namely $\epsilon_1$ and $\epsilon_2$. This over-determination allows us to use 
several criteria for the optimisation procedure and later on select the best one
that achieves results closest to experiment.\\

The optimization procedure consists of defining an objective function
and finding its minimum in the two-dimensional parameter space {$\epsilon_1$, $\epsilon_2$}.
The objective function is a quadratic consisting of the difference 
between some picked levels and those produced by the parametric potential through the
solution of the RSE with the CFM.\\

We adopted several strategies based on the following observations:
The QD value is not stable for the low levels but tends
to reach a stable value when the energy increases. 
Despite the stability of the QD for the higher levels, 
the experimental (and therefore) numerical accuracy decreases when
the energy increases.\\

Therefore a compromise should be achieved by selecting the levels in order to
define the objective function to minimize. \\

We found that a reasonable compromise should be based on the following operations that
differ with the selected rare gas: 

\begin{enumerate}
\item Pick some level and take the average of a number of higher ones.
\item Pick two high levels for which the QD has stabilised
within a given accuracy.
\item Pick a low level and a high one for which the QD has
already stabilised.
\end{enumerate}

All the above operations should yield roughly the same values for
the parameters before running the final check in order to test the
accuracy of the obtained eigenvalues.\\

\begin{table}[!ht]
\begin{center}
\begin{tabular}{c c c}
\hline
 Series & $\epsilon_1$ &  $\epsilon_2$\\
\hline   
   $l$=0 &    3.625 & 1.036\\
   $l$=1  &   3.62  & 1.06\\
   $l$=2 & 3.6344  & 1.036 \\
\hline     
\end{tabular}
\end{center}
\caption{Szydlik-Green parameters for the first three Rydberg series of Argon.}
\label{tab3}
\end{table}

The optimization program itself is based on a globally convergent
method for solving non-linear system of equations: the multidimensional
secant method developed by Broyden \cite{Broyden}.
It is based on a fast and accurate method for the iterative evaluation
of the Jacobian of the objective function needed during the minimisation 
procedure.\\

It is a Quasi-Newton method that consists of approximating the Jacobian
and updating it with an iterative procedure. It converges superlinearly
to the solution like all secant methods.\\

There are several ways to perform the integration of the RSE on the basis of the CFM. 
One may use a fixed step scheme such as the explicit RK4 method 
or a variable step (VSCA) procedure. 
Optimization wise, the RK4 method is faster than VSCA but less accurate. 
To judge the accuracy of our optimization we compare in table~\ref{tab3} 
our results for the energy levels and corresponding QD obtained 
using the RK4 and VSCA integration scheme with those of Szydlik {\sl et al.} \cite{Szydlik74}. 
The results displayed in table~\ref{tab3} clearly favor, as expected, the VSCA integration scheme. 
In contrast to the VSCA, the RK4 
integration scheme is limited to fourth order accuracy. Table \ref{tab3} further shows the sensitivity of the QD 
to the numerical values of the calculated energy levels. Despite small differences between the energy values 
obtained with the RK4 and VSCA methods, the corresponding QD's largely differ. This sensitivity of the QD
is the main motivation for using it rather than the raw energy levels in our optimization procedure.

\begin{table}[!ht]
\begin{center}
\begin{tabular}{c c c}
\hline
Experimental levels & RK4 & VSCA \\
\hline
   -0.309522&  -0.214082&  -0.310563\\
   -0.124309&   -9.87422(-2)&  -0.124506\\
    -6.76780(-2)&   -5.69743(-2)&   -6.76904(-2)\\
    -4.25540(-2)&   -3.70845(-2)&   -4.25546(-2)\\
    -2.92210(-2)&   -2.60593(-2)&   -2.92238(-2)\\
    -2.13080(-2)&   -1.93126(-2)&   -2.13062(-2)\\
    -1.62200(-2)&   -1.48846(-2)&   -1.62210(-2)\\
    -1.27620(-2)&   -1.18223(-2)&   -1.27614(-2)\\
    -1.03020(-2)&   -9.61654(-3)&   -1.03015(-2)\\
    -8.49000(-3)&   -7.97521(-3)&   -8.49011(-3)\\
    -7.11800(-3)&   -6.72092(-3)&   -7.11771(-3)\\
\hline     
\end{tabular}
\end{center}
\caption{Comparison between the experimental and calculated energy levels of the Rydberg series of Argon. 
The levels calculated with the CFM are obtained either with fixed step (RK4) or variable step (VSCA) integration. 
All values in Rydbergs.}
\label{tab4}
\end{table}

\begin{table}[!ht]
\begin{center}
\begin{tabular}{c c c}
\hline
Experimental QD & RK4 & VSCA \\
\hline
0.202561 & 0.838726 & 0.205576 \\
0.163723 & 0.817645 & 0.165967 \\
0.156063 & 0.810516 & 0.156415 \\
0.152366 & 0.807174 & 0.152400 \\
0.150046 & 0.805324 & 0.150326 \\
0.149399 & 0.804191 & 0.149110 \\
0.148104 & 0.803444 & 0.148345 \\
0.148016 & 0.802940 & 0.147808 \\
0.147664 & 0.802574 & 0.147425 \\
0.147091 & 0.802298 & 0.147161 \\
0.147199 & 0.802084 & 0.146957 \\
\hline     
\end{tabular}
\end{center}
\caption{ Comparison between the experimental and calculated QD of the Rydberg series of Argon. 
The levels calculated with the CFM are obtained either with fixed step (RK4) or variable step (VSCA) integration.}
\label{tab5}
\end{table}

\section{Vibrational spectra of  Cold Molecules}
A new kind of high precision molecular spectroscopy is probing
long-range forces between constituent atoms of molecules. This
spectroscopy is based on using light to combine two colliding
cold-trapped atoms into a tenuous molecule. 

The burgeoning field of "Photoassociation Spectroscopy" is allowing
very  precise measurement of lifetimes of the first excited states of
Alkaline atoms and observation of retardation effects and long-range
forces. It provides a means of probing accurately the weak
interaction between these atoms \cite{Jones96}.

The agreement between theory and experiment requires simultaneously a
highly accurate representation of the interaction potential as well as
a highly reliable method for the calculation of the corresponding
energy levels.

Since our aim is directed towards the latter problem, we make use of an
alternative method to evaluate the energy levels for the potential at 
hand instead of comparing to the experimental values in order to assess
the validity of our results.

The determination of the vibrational spectra of these very tenuous
molecules is extremely subtle specially for the highest levels which
play an important role in photoassociation spectroscopy. Thus a careful
control of accuracy is needed in order to diagonalise the Hamiltonian
without losing accuracy for all energies including those close to the
dissociation limit.

The magnitudes of potential energy, distance and mass values in these
kinds of molecules stand several orders of magnitude above or below what
is encountered in ordinary short-range molecules.

For instance, the typical intramolecular potential well depth at the
equilibrium distance of about 100 $a_0$ (Bohrs), is a fraction of a
cm$^{-1}$ while the reduced mass is several 10,000 electron masses. All
these extreme values require special numerical techniques in order to
avoid roundoffs, divergences, numerical instability and
ill-conditioning during processing.

\begin{figure}[htbp]
\begin{center}
\scalebox{0.9}{\includegraphics*{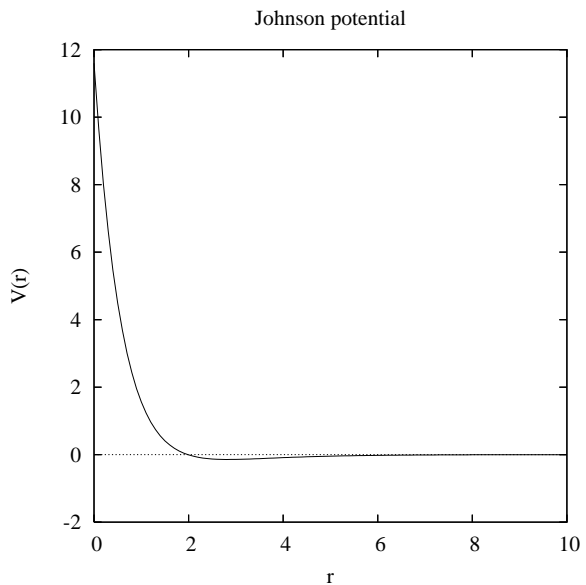}}
\end{center}
\caption{Johnson potential versus distance. The Morse parameters are $D=0.142,   a=0.815,  r_e=2.835$.
Note the shallowness and breadth of the potential minimum. We use the same units as the Morse
potential described previously in fig.~\ref{morsepot}.} 
\label{Johnson}
\end{figure}

As a first example, Johnson et al. \cite{Johnson} introduced a variant of the Morse potential
with parameters given by $D=0.142,   a=0.815,  r_e=2.835$.
While the mass parameter in the Morse case is 0.5 $m_e$, 
in the Johnson case it is very large and equal to 9114.44 $m_e$. \\
Large values of the mass parameter lead to a very shallow well
in the potential energy (see fig.~\ref{Johnson}) that means a very small
binding energy making the molecule extremely weak. This then corresponds to the so-called
long-range or cold molecule because of the extremely low molecular binding
energy. In addition the potential width around the minimum is very broad implying large 
quantum fluctuations in the ground state in contrast to the ordinarily deep
and narrow Morse potential. \\
The potential energy is given by:

\begin{equation}
        V(r)=D {[ 1  -\exp(-a \{r-r_e \}) ]}^2 -D
\end{equation}

\begin{table}[!ht]
\begin{center}
\begin{tabular}{c c c c}
\hline Index& Exact & CFM &  Johnson  \\
\hline
 0   &  497.668182   &  497.726562   &  498.    \\
 1   &  1481.66138   &  1481.67773   &  1482.    \\
 2   &  2449.65503   &  2449.69922   &  2450.    \\
 3   &  3401.64697   &  3401.70508   &  3402.    \\
 4   &  4337.6416   &  4337.69531   &  4338.    \\
 5   &  5257.63672   &  5257.68164   &  5258.    \\
 6   &  6161.63232   &  6161.66016   &  6162.    \\
 7   &  7049.62842   &  7049.63672   &  7050.    \\
 8   &  7921.62256   &  7921.67969   &  7922.    \\
 9   &  8777.62012   &  8777.70508   &  8778.    \\
 10   &  9617.61719   &  9617.67578   &  9618.    \\
 11   &  10441.6133   &  10441.6348   &  10442.    \\
 12   &  11249.6113   &  11249.6504   &  11250.    \\
 13   &  12041.6084   &  12041.6914   &  12042.    \\
 14   &  12817.6074   &  12817.6465   &  12818.    \\
 15   &  13577.6055   &  13577.6445   &  13578.    \\
 16   &  14321.6045   &  14321.666   &      \\
 17   &  15049.6045   &  15049.6191   &      \\
 18   &  15761.6035   &  15761.666   &      \\
 19   &  16457.6035   &  16457.6484   &      \\
 20   &  17137.6055   &  17137.6523   &      \\
 21   &  17801.6055   &  17801.6484   &      \\
 22   &  18449.6074   &  18449.6328   &      \\
 23   &  19081.6094   &  19081.6562   &      \\
 24   &  19697.6113   &  19697.6406   &      \\
 25   &  20297.6152   &  20297.666   &      \\
 26   &  20881.6172   &  20881.6562   &      \\
 27   &  21449.6211   &  21449.6602   &      \\
 28   &  22001.625   &  22001.668   &      \\
 29   &  22537.6289   &  22537.6328   &      \\
\hline
\end{tabular}
\caption{First thirty vibrational levels of the Johnson diatomic molecule in cm$^{-1}$. 
Johnson results (limited to the first sixteen only) are compared to the CFM and the exact values.
The corresponding analytical values are almost indistinguishable (see ref.~\cite{Johnson}). The total
number of levels is 62 with the parameters $D=0.142,   a=0.815,  r_e=2.835$.}
\label{Diatomic1}
\end{center}
\end{table}

In order to perform the eigenvalue calculation we rescale the 
parameters $D=0.142,   a=0.815,  r_e=2.835$ such that they are as close as possible
to the Morse potential described previously. The scaling yields the parameters:
$D=142.387,   a=0.815131,  r_e= 2.83459. $ These values yield in a straightforward way
all the eigenvalues (62 in total), the first thirty only being displayed in Table~\ref{Diatomic1}.

\subsection{ Vibrational energy levels of the $^{23}{\rm Na}_2$ molecule}
Since vibrational spectra of the Johnson molecule is determined accurately
and to arbitrary accuracy with the CFM, we move on to tackle the $^{23}{\rm Na}_2$ molecule. 
The energy levels of the $0^-_g$ and $1_u$ electronic states of the
$^{23}{\rm Na}_2$ molecule are determined from the Ground state up to the
continuum limit. The method is validated by comparison with previous
results obtained by Stwalley {\it et al.} \cite {Stwalley78} using the same
potential and  Trost {\it et al.} \cite {Trost98} whose
 work is based on the Lennard-Jones potential adapted to long-range molecules.

The method we use in this work adapts well to this extreme situation
with the proviso of employing a series of isospectral scaling
transformations we explain below.

We apply the CFM to the calculation of the vibrational energy levels of
a diatomic molecule where the interaction between the atoms is given by
the Movre and Pichler potential \cite {Movre77}.

We start with the $0^-_g$ electronic state of the $^{23}{\rm Na}_2$
molecule. The corresponding potential is given by:
\begin{equation}
V(r)=\frac{1}{2}[(1-3X)+\sqrt{1-6X+81X^2}]
\end{equation}

where $X=\frac{C(0^-_g)}{9r^3\Delta}$.  \\
$r$ is the internuclear distance and the parameter $C(0^-_g)$ is such that:
\begin{equation}
\lim_{r \rightarrow +\infty} V(r) \rightarrow -\frac{C_3(0^-_g)}{r^3}
\end{equation}
Identification of the large $r$ limit yields the result:
$C_3(0^-_g)=C(0^-_g)/3$. In the calculations below, we have used
$C_3(0^-_g)=6.390$ Hartrees.$a_0^3$ like Stwalley {\it et al.}\cite{Stwalley78}. The
parameter $\Delta=1.56512.10^{-4}$ Rydbergs is the atomic spin-orbit
splitting. Given $C_3(0^-_g)$ and $\Delta$, the equilibrium
internuclear distance is $r_e=71.6 a_0$.

\begin{figure}[htbp]
\begin{center}
\scalebox{0.5}{\includegraphics*{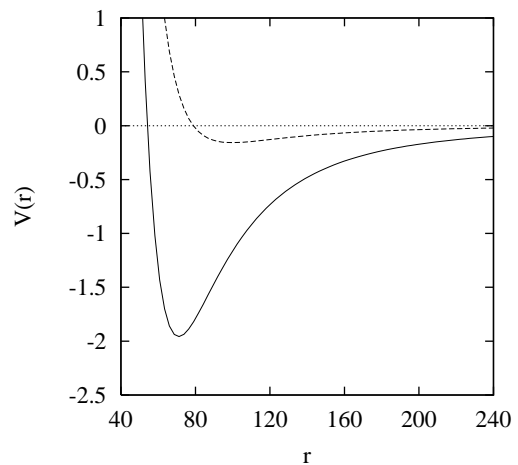}}
\end{center}
\caption{Potential energy in cm$^{-1}$ for the $0^-_g$ and $1_u$ electronic states 
(upper shallow curve) of the $^{23}{\rm Na}_2$ molecule. The radial distance r is in Bohrs ($a_0$) units.}
\label{potNa}
\end{figure}

The peculiarities of the  $0^-_g$ and $1_u$ electronic states of the $^{23}{\rm Na}_2$ molecule
are evident in fig.~\ref{potNa}.
The $1_u$  state is higher that the $0^-_g$ and the number of vibrational
levels is smaller because the potential is shallower. This has important consequences as we
we explore the vibrational spectrum below.

\begin{figure}[htbp]
\begin{center}
\scalebox{0.5}{\includegraphics*{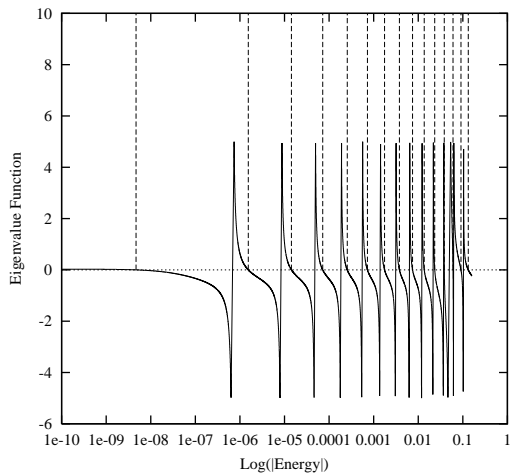}}
\end{center}
\caption{Behavior of the eigenvalue function $F(E)$ with energy on a
semi-log scale for the $1_u$ electronic state of the $^{23}{\rm Na}_2$
molecule. The vertical lines indicate the eigenvalue position. Energies
are in cm$^{-1}$.} \label{fig3}
\end{figure}

We scale all energies with a factor $E_0$ (usually cm$^{-1}$) with the
use of equation (1). Then we scale all distances with a typical length
$L_0$ transforming the RSE appropriately. This double transformation is
reflected generally in the potential coefficients preserving thus the
functional form of the potential. 

In order to gauge the accuracy of the spectra, we perform the
integration of the RSE with two different methods: the fixed step
RK4 and the variable step VSCA methods \cite {Kobeissi91}. 

The VSCA method is based on a series expansion of the potential and the
corresponding solution to an order such that a required tolerance
criterion is met. Ideally, the series coefficients are determined
analytically to any order, otherwise loss of accuracy occurs leading
quickly to numerical uncertainties as discussed later.

\begin{figure}[htbp]
\begin{center}
\scalebox{0.5}{\includegraphics*{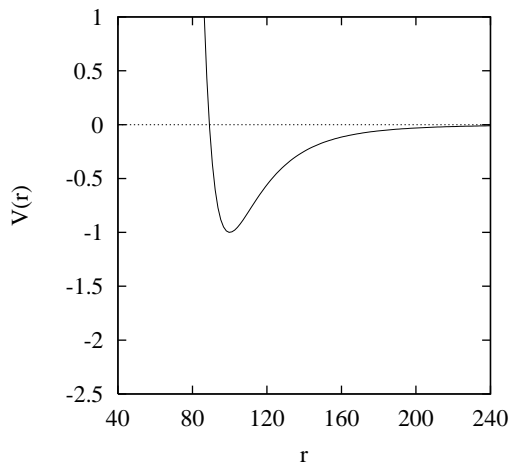}}
\end{center}
\caption{Potential energy of the Trost {\it et al.} \cite{Trost98} model.
The energy is in $\epsilon$ units where $\epsilon$ is the potential
well depth and the radial distance is in $\frac{a_0}{\sqrt{B}}$ where
$B$ is a scaled mass. We use the same interval for the potential
and radial distance in order to ease the comparison with fig.~\ref{potNa} 
pertaining to the $^{23}{\rm Na}_2$ molecule.} \label{fig4}
\end{figure}

Table \ref{0g} shows the results we obtain with the RK4 method.
The limitation of the RK4 method to fourth order hampers the finding of
levels beyond the 33rd (see table~\ref{0g}). In order to find the higher
levels we have to select an algorithm that enables us to tune the
accuracy well beyond the fourth order.

Pushing the accuracy within the framework of a fixed step method has
the effect of reducing substantially the integration step. In order to
avoid this problem, we use a variable step that adjusts itself to the
desired accuracy, the VSCA method.

This method is powerful and flexible enough to find all the desired energy
levels and allows us to find one additional level that was not detected before. 
It should be noted that the last three levels given by Stwalley at al.
\cite{Stwalley78} were extrapolated and not calculated. The agreement
between our calculated levels and those of Stwalley {\it et al.}
\cite{Stwalley78} is quite good. We believe that the small
discrepancy, increasing as we progress towards the dissociation limit,
is due to a loss of accuracy associated with traditional methods in
sharp contrast with the CFM.

\begin{table}[!ht]
\begin{center}
\begin{tabular}{ c c c c }
\hline Index&  RK4 (cm$^{-1}$)&  Stwalley {\it et al.} (cm$^{-1}$)&
Ratio  \\ \hline
  1&    -1.7864563&   -1.7887&    1.00126\\
   2&   -1.5595812&   -1.5617&    1.00136\\
   3&   -1.3546211&   -1.3566&    1.00146\\
   4&   -1.1704091&   -1.1723&    1.00162\\
   5&   -1.0057168&   -1.0075&    1.00177\\
   6&   -0.8592746&   -0.86087&    1.00186\\
   7&   -0.7297888&   -0.73125&    1.00200\\
   8&   -0.6159592&   -0.61729&    1.00216\\
   9&   -0.5164938&   -0.51770&    1.00234\\
   10&  -0.4301231&   -0.43120&    1.00250\\
   11&  -0.3556114&   -0.35657&    1.00270\\
   12&  -0.2917683&   -0.29261&    1.00288\\
   13&  -0.2374567&   -0.23820&    1.00313\\
   14&  -0.1916004&   -0.19224&              1.00334\\
   15&  -0.1531898&   -0.15374&              1.00359\\
   16&  -0.1212858&        -0.12176&         1.00391\\
   17&  -9.5022481(-02)&    -9.5438(-02)&    1.00437\\
   18&  -7.3608067(-02)&    -7.3940(-02)&    1.00451\\
   19&  -5.6325397(-02)&    -5.6599(-02)&    1.00486\\
   20&  -4.2530440(-02)&    -4.2754(-02)&    1.00526\\
   21&  -3.1650256(-02)&    -3.1831(-02)&    1.00571\\
   22&  -2.3180032(-02)&    -2.3323(-02)&    1.00617\\
   23&  -1.6679434(-02)&    -1.6791(-02)&    1.00669\\
   24&  -1.1768423(-02)&    -1.1854(-02)&    1.00727\\
   25&  -8.1226859(-03)&    -8.1873(-03)&    1.00795\\
   26&  -5.4687973(-03)&    -5.5165(-03)&    1.00872\\
   27&  -3.5792655(-03)&    -3.6136(-03)&    1.00959\\
   28&  -2.2675456(-03)&    -2.2916(-03)&    1.01061\\
   29&  -1.3831324(-03)&    -1.3995(-03)&    1.01183\\
   30&  -8.0680818(-04)&    -8.1747(-04)&    1.01321\\
   31&  -4.4611287(-04)&    -4.5276(-04)&    1.01490\\
   32&  -2.3077899(-04)&    -2.3503(-04)&    1.01842\\
   33&  -9.4777816(-05)&    -1.1252(-04)&    1.18720\\
   34& 	                   &-4.8564(-05)&               \\
   35& 		           &-1.8262(-05)&               \\
   36&	 	           &-5.6648(-06)&               \\
   37& 		           &-1.3175(-06)&             \\
   38&	 	           &-1.9247(-07)&               \\
   39& 		           &-1.1215(-08)&             \\
   40&	 	           &-4.1916(-11)&              \\
\hline
\end{tabular}
\caption{Vibrational levels for the $0^-_g$ electronic state of the
$^{23}{\rm Na}_2$ molecule as obtained with a fixed step RK4 method,
Stwalley {\it et al.} \cite{Stwalley78} results and the corresponding ratio.
Levels 34-40 were not found by the RK4 method due to the precision
limited to fourth order.} 
\label{0g}
\end{center}
\end{table}

\begin{table}[!ht]
\begin{center}
\begin{tabular}{ c c c c }
\hline Index& Stwalley {\it et al.} (cm$^{-1}$) & VSCA (cm$^{-1}$) & Ratio \\
\hline
  1&   -1.7887&   -1.7864488&             1.00126\\
   2&  -1.5617&   -1.5595638&             1.00137\\
   3&  -1.3566&   -1.3546072&             1.00147\\
   4&  -1.1723&   -1.1703990&             1.00162\\
   5&  -1.0075&   -1.0057071&              1.00178\\
   6&  -0.86087&   -0.8592631&             1.00187\\
   7&  -0.73125&   -0.7297908&             1.00200\\
   8&  -0.61729&   -0.6159534&             1.00202\\
   9&  -0.51770&   -0.5164882&             1.00235\\
  10&  -0.43120&   -0.4301217&             1.00251\\
  11&  -0.35657&   -0.3556148&             1.00269\\
  12&  -0.29261&   -0.2917693&             1.00288\\
  13&  -0.23820&   -0.2374560&             1.00313\\
  14&  -0.19224&   -0.1916002&             1.00334\\
  15&  -0.15374&   -0.1531893&             1.00359\\
  16&  -0.12176&         -0.1212854&       1.00391\\
  17&  -9.5438(-02)&     -9.5022588(-02)&   1.00437\\
  18&  -7.3940(-02)&     -7.3608452(-02)&   1.00450\\
  19&  -5.6599(-02)&     -5.6325744(-02)&   1.00485\\
  20&  -4.2754(-02)&     -4.2530867(-02)&   1.00525\\
  21&  -3.1831(-02)&     -3.1650591(-02)&   1.00570\\
  22&  -2.3323(-02)&     -2.3180420(-02)&   1.00615\\
  23&  -1.6791(-02)&     -1.6679756(-02)&   1.00667\\
  24&  -1.1854(-02)&     -1.1768655(-02)&   1.00725\\
  25&  -8.1873(-03)&     -8.1228816(-03)&   1.00793\\
  26&  -5.5165(-03)&     -5.4689541(-03)&   1.00869\\
  27&  -3.6136(-03)&     -3.5793742(-03)&   1.00956\\
  28&  -2.2916(-03)&     -2.2676168(-03)&   1.01058\\
  29&  -1.3995(-03)&     -1.3831806(-03)&   1.01180\\
  30&  -8.1747(-04)&     -8.0683859(-04)&   1.01318\\
  31&  -4.5276(-04)&     -4.4613249(-04)&  1.01486\\
  32&  -2.3503(-04)&     -2.3110168(-04)&   1.01700\\
  33&  -1.1252(-04)&     -1.1035443(-04)&   1.01962\\
  34&  -4.8564(-05)&     -4.7468345(-05)&   1.02308\\
  35&  -1.8262(-05)&     -1.7767388(-05)&   1.02784\\
  36&  -5.6648(-06)&     -5.4747950(-06)&   1.03471\\
  37&  -1.3175(-06)&     -1.2597092(-06)&   1.04588\\
  38&  -1.9247(-07)&     -1.2716754(-07)&   1.51352\\
  39&  -1.1215(-08)&                                     \\
  40&  -4.1916(-11)&                                      \\
\hline
\end{tabular}
\caption{Vibrational levels for the $0^-_g$ electronic state of the
$^{23}{\rm Na}_2$ molecule as obtained with Stwalley {\it et al.} results, the
VSCA method and the
corresponding ratio. Levels 38, 39 and 40 of Stwalley {\it et al.}
\cite{Stwalley78} are extrapolated with LeRoy and Bernstein
\cite{LeRoy70} semi-classical formulae.} 
\label{stwg}
\end{center}
\end{table}

\begin{table}[!ht]
\begin{center}
\begin{tabular}{ c c c c }
\hline Index&  RK4 (cm$^{-1}$) &  Stwalley {\it et al.}(cm$^{-1}$) &
Ratio \\ \hline
   1&    0.1319536     &    0.13212&        1.00126\\
   2&    9.0057392(-02)&    9.0192(-02)&    1.00149\\
   3&    5.9472159(-02)&    5.9574(-02)&    1.00171\\
   4&    3.7821597(-02)&    3.7896(-02)&    1.00197\\
   5&    2.3027828(-02)&    2.3080(-02)&    1.00227\\
   6&    1.3324091(-02)&    1.3359(-02)&    1.00262\\
   7&    7.2562182(-03)&    7.2787(-03)&    1.00310\\
   8&    3.6714971(-03)&    3.6849(-03)&    1.00365\\
   9&    1.6950002(-03)&    1.7024(-03)&    1.00437\\
   10&   6.9533077(-04)&    6.9904(-04)&    1.00533\\
   11&   2.4102039(-04)&    2.4492(-04)&    1.01618\\
   12&                 &    6.8430(-05)&           \\
   13&                 &    1.3446(-05)&           \\
   14&                 &    1.4122(-06)&           \\
   15&                 &    3.8739(-08) &         \\
   16&                 &    1.2735(-12) &            \\

\hline
\end{tabular}
\caption{Vibrational levels for the $1_u$ electronic state of the
$^{23}{\rm Na}_2$ molecule as obtained with the RK4 method, Stwalley
{\it et al.} results and the corresponding ratio. RK4 found only 11 levels
and levels 15 and 16 of Stwalley {\it et al.} are found by extrapolation.}
\label{1u}
\end{center}
\end{table}

\begin{table}[!ht]
\begin{center}
\begin{tabular}{ c c c c }
\hline Index& Stwalley {\it et al.} (cm$^{-1}$) & VSCA (cm$^{-1}$) & Ratio \\
\hline
  1&     0.13212&        0.13244150&         1.00243\\
   2&    9.0192(-02)&    9.07598688(-02)&    1.00630\\
   3&    5.9574(-02)&    6.01645742(-02)&    1.00991\\
   4&    3.7896(-02)&    3.83982868(-02)&    1.01325\\
   5&    2.3080(-02)&    2.34584275(-02)&    1.01640\\
   6&    1.3359(-02)&    1.36187753(-02)&    1.01944\\
   7&    7.2787(-03)&    7.44243743(-03)&    1.02250\\
   8&    3.6849(-03)&    3.77999552(-03)&    1.02581\\
   9&    1.7024(-03)&    1.75281307(-03)&    1.02961\\
   10&   6.9904(-04)&    7.23013793(-04)&    1.03430\\
   11&   2.4492(-04)&    2.54856605(-04)&    1.04057\\
   12&   6.8430(-05)&    7.18249908(-05)&    1.04961\\
   13&   1.3446(-05)&    1.43120199(-05)&    1.06441\\
   14&   1.4122(-06)&    1.53931042(-06)&    1.09001\\
   15&   3.8739(-08)&    4.66022073(-09)&            \\
   16&   1.2735(-12)&                                \\
\hline
\end{tabular}
\caption{Vibrational levels for the $1_u$ electronic state of the
$^{23}{\rm Na}_2$ molecule as obtained by Stwalley {\it et al.}, the 
VSCA and the corresponding ratio. A
new 15th level is obtained with the VSCA method.} 
\label{stwu}
\end{center}
\end{table}

The estimation of accuracy of the results hinges basically
on two operations, integration and determination of the zeroes
of the eigenvalue function $F(E)$. The superiority of the VSCA
method is observed in the determination of the upper levels that are 
not detected by the RK4 method (see Tables 1 and 2).
In addition, it is observed in the behavior of the eigenvalue ratio
versus the index. While in both cases (RK4 and VSCA) the ratio increases steadily as
the index increases because we are probing higher excited states, in the 
RK4 case it rather blows up as dissociation is approached. We use typically
series expansion to order 12 in VSCA with a tolerance of $10^{-8}$. 
In the root search of $F(E)$, the tolerance required for a zero
to be considered as an eigenvalue is $10^{-15}$.
This does not imply that we disagree as strongly as 0.13\%, for instance,
with the Ground state value (see Tables 1 and 2) found by Stwalley {\it et al.}
\cite {Stwalley78} for the simple reason, we use a splitting energy 
$\Delta=1.56512.10^{-4}$ Rydbergs  corresponding to an equilibrium
internuclear distance $r_e=71.6 a_0$. 
Stwalley {\it et al.} do not provide explicitly the value of $\Delta$ they use, 
however Jones {\it et al.} \cite{Jones96} use a value that is slightly
different.\\

We move on to treat the $1_u$ electronic state of the $^{23}{\rm Na}_2$ molecule.
The potential associated with the $1_u$ electronic state of the
$^{23}{\rm Na}_2$ molecule is a lot more involved. Its VSCA implementation
is particularly difficult because of the complex functional of the
potential as we explain below. Analytically, the VSCA algorithm
requires performing a Taylor series expansion to any order around an
arbitrary point \cite{Kobeissi90}.

Numerically, this is still an open problem for arbitrary functions.
If one turns toward the use of LISP based symbolic manipulation 
techniques cumbersome expressions are produced hampering any progress.
Special methods based on analytical fitting
expressions are needed in order to turn the series coefficients into a
more manageable form (see refs~\cite{Fornberg1,Fornberg2}).

The first step is to determine the $1_u$ electronic state of the
$^{23}{\rm Na}_2$ molecule by solving the Movre {\it et al.} \cite {Movre77}
secular equation such that:
\begin{equation}
V(r)=\Delta[-2\sqrt{Q}\cos(\frac{\theta-2\pi}{3})-\frac{a}{3}-1]
\end{equation}
where $a=-2-6X$ and $X=\frac{C(1_u)}{9r^3\Delta}$. In addition,
$\theta=\cos^{-1}( \frac{1+270X^3}{\sqrt{(1+63X^2)^3}})$, and
$Q=\frac{1+63X^2}{9}$.

The parameter $C(1_u)$ is such that:
\begin{equation}
\lim_{r \rightarrow +\infty} V(r) \rightarrow -\frac{C_3(1_u)}{r^3}
\end{equation}

Identification of the large $r$ limit yields to the result:
$C_3(1_u)=C(1_u)(\sqrt{7}-2)/9$.
 We have used in the calculations below  $C_3(1_u)$=1.383 Hartrees.$a_0^3$ like
Stwalley {\it et al.} The parameter $\Delta=1.56512.10^{-4}$ Rydbergs is the
same as for the $0^-_g$ state.

Table \ref{tab3} displays the results we obtain with the RK4
method that cannot find more than 11 levels due to accuracy limitations.

\begin{figure}[htbp]
\begin{center}
\scalebox{0.5}{\includegraphics*{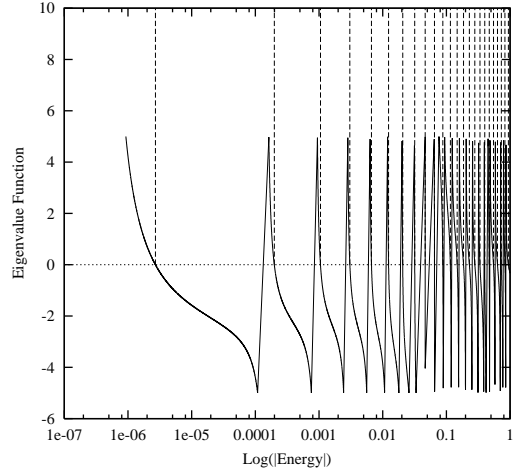}}
\end{center}
\caption{ Behavior of the eigenvalue function $F(E)$ with energy on a
semi-log scale for the Trost {\it et al.} \cite{Trost98} Lennard-Jones
molecule. The vertical lines indicate the eigenvalue position. Energies
are in potential depth $\epsilon$ units.} \label{fig5}
\end{figure}

The next results for the $1_u$ electronic state of the $^{23}{\rm Na}_2$
molecule are obtained with the VSCA method as shown in  table~\ref{tab4}. 
We find an additional 15-th level in contrast to  Stwalley {\it et al.} who found
fourteen and extrapolated the last two levels.

The corresponding graph of the eigenvalue function is displayed in
Fig.~\ \ref{fig3} below.\\

\subsection{Lennard-Jones molecules}
We apply our methodology to the Lennard-Jones case. Our results are
compared to the results obtained by Trost {\it et al.} \cite{Trost98}. We
start with the levels obtained with the RK4 method. The energy unit is
$\epsilon$ the absolute value of potential well depth. The Asymmetric
Lennard-Jones potential (ALJ) is given by:

\begin{equation}
V(r)= {\frac{C_1}{r^\beta}} - {\frac{C_2}{r^\alpha}}
\end{equation}

It depends on $C_1$ and $C_2$ that yield
an equilibrium distance at $r=r_{e}$  and a potential depth 
$-\epsilon$. In order to compare with the appropriate litterature,
Trost {\it et al.} \cite{Trost98} use rather the general parameterisation:
\begin{equation}
C_1 = \frac{\epsilon}{(\beta-\alpha)}\alpha r_{e}^\beta
\hspace{0.1cm}, \hspace{0.1cm} C_2=\frac{\epsilon}{(\beta-\alpha)}\beta
r_{e}^\alpha
\end{equation}
It is scaled in such a way that the energy is expressed in units of the
potential well depth $-\epsilon$. When $\alpha=6$, $\beta=12$ we obtain
$r_{e}=\sqrt[6]{\frac{2C_{1}}{C_2}}$ and $\epsilon=
\frac{C_2^2}{4C_{1}}$. and the radial distance is in
$\frac{a_0}{\sqrt{B}}$ where $B$ is a reduced scaled mass given by $B=2
\mu \epsilon {r^2_{e}}$. Numerically Trost {\it et al.} use $B=10^4$ which
is the order of magnitude encountered in long-range molecules within
the framework of their system of units.
The potential energy in these units is displayed in Fig.~\ \ref{fig4}.
The RK4 methods yields the results displayed in Table \ref{tab5}. \\

The accuracy limitation of the RK4 results in losing the uppermost
level 24. Thus we move on to the results obtained with the superior
VSCA method.
All levels are obtained with the VSCA and the agreement with Trost et
al. results is perfect as witnessed by the ratio values of Table 6.
The eigenvalue function obtained with the VSCA method as a function of
energy is displayed in Fig.~\ \ref{fig5}.

\section{Test of Bohr Correspondance Principle}

LeRoy and Bernstein ~\cite{LeRoy70} used a WKB approximation to derive a semi-classical
 formula (SCF) for the distribution
of vibrational levels near the dissociation limit of diatomic molecules. If the long-range
interatomic potential is of the form $D-C_n/r^n$ their formula allows the determination of 
 the dissociation energy $D$ as well as the tail of the potential ($n$ and $C_n$) from the 
 experimental energies of the highest vibrational levels. Furthermore, this analysis opens 
 the possibility of testing directly Bohr Correspondence Principle (BCP) that states 
 the agreement between semi-classical
and quantum results improves for larger quantum numbers.\\ 
  
We consider the rotationless (L=0) $0^-_g$ and the $1_u$ electronic states of the $^{23}{\rm Na}_2$ molecule 
correlating asymptotically to the atomic $^{2}{P}_{3/2}$ and $^{2}{S}_{1/2}$ states. The interaction between the
atoms is represented by the Movre-Pichler ~\cite{Movre77} potential that behaves as $-1/r^3$ as 
$r\rightarrow +\infty$ and we use the same parameters as those of Stwalley {\it et al.} ~\cite{Stwalley78}. 
The Asymmetric Lennard-Jones potential (ALJ) $C_1 r^{-\beta}-C_2 r^{-\alpha}$ is also considered in
 the highly unsymmetric case $C_2 >> C_1$ (we tackle here the case $\beta=12$ and $\alpha=6$, 
 therefore we have asymptotically $-1/r^6$) and the Trost {\it et al.} ~\cite {Trost98} case. 
 Both potentials are tailored to the study of Long-range molecules.\\

The determination of the vibrational spectra of these very tenuous molecules
is extremely subtle especially for the highest levels which play an important role in photoassociation 
spectroscopy. Special care is needed in order to diagonalise the Hamiltonian without losing 
accuracy for all energies including those close to the dissociation limit.\\

The magnitudes of potential energy, distance and mass values in this kind of molecules stand several 
orders of magnitude above or below what is encountered in ordinary short-range molecules. For instance the typical intramolecular 
potential well depth at the equilibrium distance of about 100 $a_0$ (Bohrs), is a 
fraction of a cm$^{-1}$ while the reduced mass is several 10,000 electron masses as 
observed in the Johnson case previously. All these
extreme values require special numerical techniques in order to avoid roundoffs, divergences, 
numerical instability and ill-conditioning during processing.\\

Accuracy and its control being of paramount importance in this work,  
the CFM~\cite {Kobeissi82} is an excellent candidate because it bypasses 
the calculation of the eigenfunctions. This avoids losing accuracy associated
 with the numerical calculation specially with rapidly oscillating wave functions of highly excited states close to
the dissociation limit. 

The semi-classical approximation has been already discussed in the literature ~\cite {Boisseau98, Trost98}.  
Here we refer to the original LeRoy and Bernstein formulation.

The next section is a discussion of the 
the Semi-classical analysis as formulated by LeRoy and Berstein with
its implications. We present further
the results we obtain for the vibrational levels of the $^{23}{\rm Na}_2$ molecule  
$0^-_g$ and $1_u$ electronic states. We present an additional validation of the method with the ALJ  molecular potential.

\subsection{Semi-classical analysis}
The derivation of the LeRoy and Bernstein ~\cite{LeRoy70} SCF is based on the WKB condition
for the eigenvalues of a potential $V(r)$:
\begin{equation}
v+1/2=\frac{\sqrt{2 \mu}}{\pi \hbar}\int_{R_1(v)}^{R_2(v)} \sqrt{E(v)-V(R)} \, dR
\label{e.1}
\end{equation} 
where $\mu$ is the reduced mass, $E(v)$ is the energy of the level indexed by $v$ and 
$R_1(v), R_2(v)$ are the the corresponding classical turning points.\\

Near the dissociation limit, one considers $E(v)$ as a continuous function of $v$ and
by differentiation of eq.~(\ref{e.1}) one obtains:
\begin{equation}
\frac{dv}{dE(v)}=\frac{\sqrt{ \mu /2}}{\pi \hbar}\int_{R_1(v)}^{R_2(v)} \frac{1}{\sqrt{E(v)-V(R)}} \, dR
\end{equation}

Using the asymptotic expression of $V(r)=D-C_n/r^n$ and changing the variable of integration to
$y=R_2(v)/R$, we obtain:
\begin{equation}
\frac{dv}{dE(v)}=\frac{\sqrt{ \mu /2}}{\pi \hbar}\int_{1}^{R_2(v)/R_1(v)} y^{-2} {(y^n-1)}^{-1/2} \, dy
\end{equation}

This integral can be expressed in terms of incomplete Euler Beta functions we use to check the accuracy 
and validate the results. LeRoy and Bernstein ~\cite{LeRoy70} SCF is derived by taking the 
limit ${R_2(v)/R_1(v) \rightarrow +\infty}$ which leads to:
\begin{equation}
\frac{dE(v)}{dv}=K_n {[D-E(v)]}^{\frac{(n+2)}{2n}}
\label{e.4}
\end{equation}

\begin{table}[!ht]
\begin{center}
\begin{tabular}{ c c c c }
\hline Index& RK4 &  Trost {\it et al.} & Ratio \\ \hline
   1&   -0.9410450&      -0.9410460&    1.000001\\
   2&   -0.8299980&      -0.8300020&    1.000005\\
   3&   -0.7276400&      -0.7276457&    1.000008\\
   4&   -0.6336860&      -0.6336930&    1.000011\\
   5&   -0.5478430&      -0.5478520&    1.000017\\
   6&   -0.4698130&      -0.4698229&    1.000021\\
   7&   -0.3992870&      -0.3992968&    1.000025\\
   8&   -0.3359470&      -0.3359561&    1.000027\\
   9&   -0.2794670&      -0.2794734&           1.000023\\
  10&   -0.2295070&      -0.2295117&           1.000021\\
  11&   -0.1857220&      -0.1857237&           1.000009\\
  12&   -0.1477510&      -0.1477514&           1.000003\\
  13&   -0.1152270&      -0.1152259&           0.999990\\
  14&   -8.776970(-02)&  -8.7766914(-02)&      0.999968\\
  15&   -6.498640(-02)&  -6.4982730(-02)&      0.999944\\
  16&   -4.647400(-02)&  -4.6469911(-02)&      0.999912\\
  17&   -3.181750(-02)&  -3.1813309(-02)&      0.999868\\
  18&   -2.059000(-02)&  -2.0586161(-02)&      0.999814\\
  19&   -1.235370(-02)&  -1.2350373(-02)&      0.999731\\
  20&   -6.659580(-03)&  -6.6570240(-03)&      0.999616\\
  21&   -3.048890(-03)&  -3.0471360(-03)&      0.999425\\
  22&   -1.053690(-03)&  -1.0527480(-03)&      0.999106\\
  23&   -1.645210(-04)&  -1.9834000(-04)&      1.205560\\
  24&                 &  -2.6970000(-06) &              \\

\hline
\end{tabular}
\caption{Quantum levels of a Lennard-Jones molecule in $\epsilon$
units, the depth of the potential well as obtained by the RK4 method,
Trost {\it et al.} and the corresponding ratio. Only the last 24th level was
missed by the RK4 method.} 
\label{ljt}
\end{center}
\end{table}

\begin{table}[!ht]
\begin{center}
\begin{tabular}{ c c c c }
\hline Index& VSCA &  Trost {\it et al.} & Ratio \\ \hline
 1&    -0.9410443&        -0.9410460&        1.000002\\
   2&  -0.8299963&        -0.8300020&        1.000007\\
   3&  -0.7276415&        -0.7276457&        1.000006\\
   4&  -0.6336915&        -0.6336930&        1.000002\\
   5&  -0.5478480&        -0.5478520&        1.000007\\
   6&  -0.4698206&        -0.4698229&        1.000005\\
   7&  -0.3992947&        -0.3992968&        1.000005\\
   8&  -0.3359533&        -0.3359561&        1.000008\\
   9&  -0.2794718&        -0.2794734&        1.000005\\
  10&  -0.2295109&        -0.2295117&        1.000003\\
  11&  -0.1857222&        -0.1857237&        1.000008\\
  12&  -0.1477498&        -0.1477514&        1.000010\\
  13&  -0.1152247&        -0.1152259&        1.000010\\
  14&  -8.7766358(-02)&   -8.7766914(-02)&   1.000006\\
  15&  -6.4982534(-02)&   -6.4982730(-02)&   1.000003\\
  16&  -4.6469838(-02)&   -4.6469911(-02)&   1.000002\\
  17&  -3.1813146(-02)&   -3.1813309(-02)&   1.000005\\
  18&  -2.0585953(-02)&   -2.0586161(-02)&   1.000010\\
  19&  -1.2350173(-02)&   -1.2350373(-02)&   1.000016\\
  20&  -6.6568735(-03)&   -6.6570240(-03)&   1.000023\\
  21&  -3.0470500(-03)&   -3.0471360(-03)&   1.000028\\
  22&  -1.0526883(-03)&   -1.0527480(-03)&   1.000057\\
  23&  -1.9832170(-04)&   -1.9834000(-04)&   1.000092\\
  24&  -2.6957891(-06)&   -2.6970000(-06)&   1.000449\\
\hline
\end{tabular}
\caption{Quantum levels of a Lennard-Jones molecule in $\epsilon$
units, the depth of the potential well as obtained by the VSCA method,
Trost {\it {\it et al.}} and the corresponding ratio.} 
\label{trostt}
\end{center}
\end{table}

\begin{figure}[htbp]
\begin{center}
\scalebox{0.5}{\includegraphics*{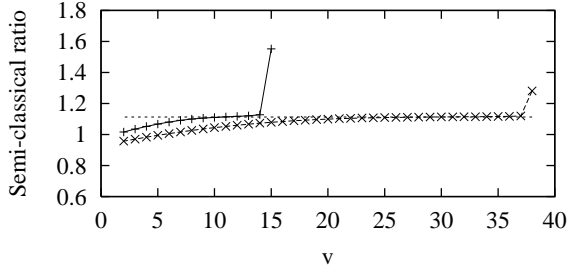}}
\end{center}
\caption{Approach of the semi-classical limit for the $0^-_g$ and $1_u$ (upper curve) electronic states 
of the $^{23}{\rm Na}_2$ molecule with the vibrational quantum  number.}
\label{diff3}
\end{figure}

where $K_n$ is a constant we calculate below for n=3 and n=6. Integrating the above equation yields:
\begin{equation}
 {[D-E(v)]}^{1/2-1/n}=\frac{(n+2)}{2n} K_n (v_{max}-v+v_D)
\label{e.5}
\end{equation}

where $v_{max}$ is the uppermost vibrational level index and $v_D$  is generally not
an integer that depends on interactions having shorter range ~\cite{LeRoy70, Stwalley78}.

Taking the origin of energies at the dissociation limit and following Gao ~\cite{Gao99}, we define a length scale with:
\begin{equation}
\beta_n=({\frac{2\mu C_n}{\hbar^2}})^{\frac{1}{(n-2)}}
\end{equation}

This length should be larger than any length scale encountered in the system.
In the case n=3, one defines a dimensionless bound-state energy by the scaling:

\begin{equation}
\epsilon_s(v)=\frac{1}{4} \frac{E(v)}{(\hbar^2/2 \mu)
 {(1/\beta_3)}^2 }
\end{equation}

With the use of  eq.~(\ref{e.5}) this leads to:
\begin{equation}
{[-\epsilon_s(v)]}^{1/6}=\frac{{[\Gamma(1/3)]}^3} {2^{5/3} 3^{1/2} \pi} (v_{max}-v+v_D)
\label{e.6}
\end{equation}

This implies that the spacing between two scaled neighbouring 
vibrational levels taken to power (1/6) is a universal constant given by 
$\frac{{[\Gamma(1/3)]}^3} {2^{5/3} 3^{1/2} \pi}$ numerically equal to  1.11292.

In the Lennard-Jones and the Trost {\it et al.} cases (n=6), the scaled bound-state energy takes the form:

\begin{equation}
\epsilon_s(v)=\frac{1}{16} \frac{E(v)}{({\hbar^2}/{2 \mu})  {(1/\beta_6)}^2 }
\end{equation}

Proceeding like the previous case, eq.~(\ref{e.5}) leads to:
\begin{equation}
{[-\epsilon_s(v)]}^{1/3}=\frac{{[\Gamma(1/3)]}^3} {2^{5/3} \pi} (v_{max}-v+v_D)
\end{equation}

In this case, the spacing between two scaled neighbouring vibrational levels taken to power (1/3) is given by
$\frac{{[\Gamma(1/3)]}^3} {2^{5/3} \pi}$ numerically equal to 1.92763. Thus, one has to determine 
the vibrational levels and examine the limit of $E(v)$ as $v$ increases.

\subsection{Test with the $^{23}{\rm Na}_2$ molecule}
We apply the CFM to the calculation of the vibrational energy levels of a 
diatomic molecule where the interaction between the atoms is given by the Movre and Pichler 
potential ~\cite {Movre77, Jones96}.\\

After scaling all energies with a factor $E_0$ (usually cm$^{-1}$) and all distances with a typical
length $L_0$  the potential coefficients in the RSE are transformed in a way such that
the functional form of the potential is preserved. \\

\begin{widetext}

\begin{figure}[h!]
\centering
\scalebox{0.6}{\includegraphics[angle=-90]{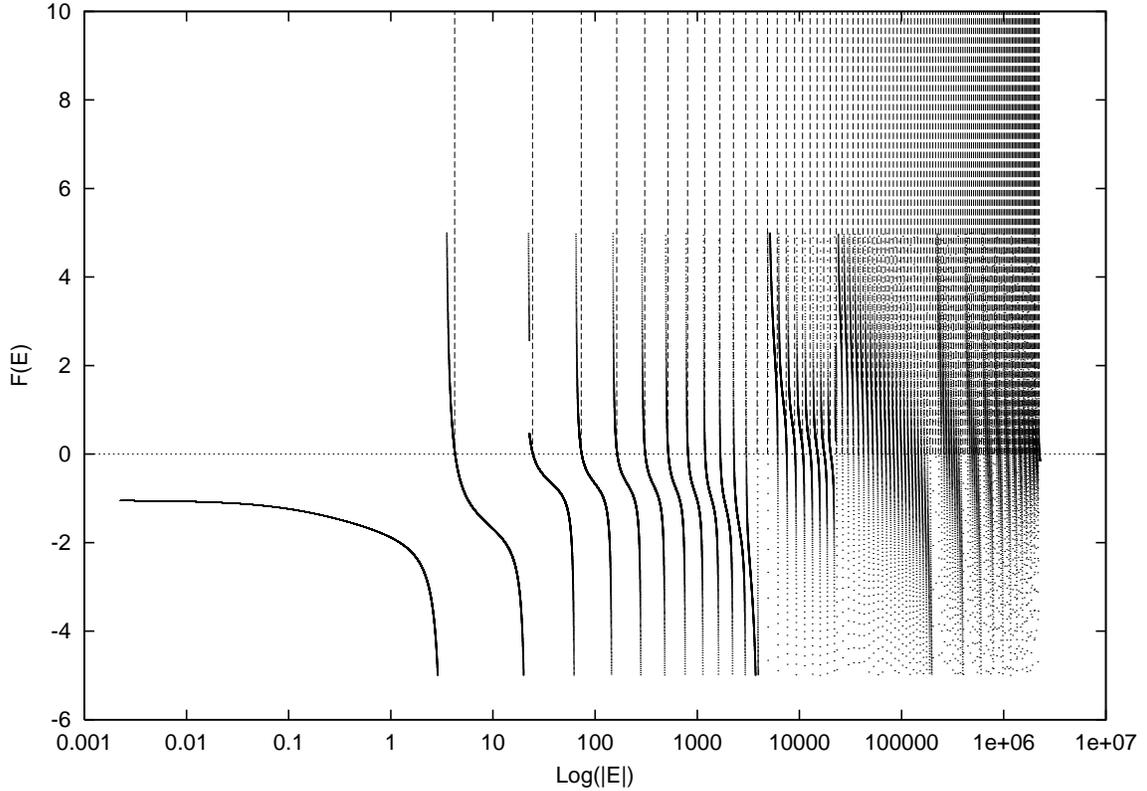}}
\caption{Behavior of the eigenvalue function $F(E)$ with energy on a
semi-log scale for the Trost {\it et al.} \cite{Trost98}  ALJ $(C_1=1, C_2=3000)$ 
molecule. The vertical lines indicate the eigenvalue position. Energies
are in potential depth $\epsilon$ units.} 
\label{LJ6}
\end{figure}

\end{widetext}

The eigenvalue function $F(E)$ depicted in fig.~\ref{LJ6} shows that the $\tan(E)$ shape is 
observed again indicating that the CFM is able to perform accurately as previously
despite the extreme sensitivity of this problem.

The integration of the RSE is performed with the VSCA method
 ~\cite {Kobeissi91}. This method is superior to fixed step methods and is based on a series
 expansion of the potential and the corresponding solution to an order such that a required 
tolerance criterion is met. Ideally, the series coefficients are determined analytically to
 any order, otherwise loss of accuracy occurs leading quickly to numerical uncertainties.\\

This method is powerful and flexible enough to find all the sought levels  plus additional 
levels not detected before by Stwalley {\it et al.} ~\cite{Stwalley78}. It should be noted that 
the last three levels given by 
Stwalley at al. ~\cite{Stwalley78} were extrapolated and not calculated. The agreement 
between the rest of our own calculated levels and those of Stwalley {\it et al.} is quite good.\\

We find all sought  levels  plus an  additional level that was not detected before. 
The agreement between our calculated levels and those of Stwalley {\it et al.} ~\cite{Stwalley78} 
turns out to be very good.\\

The semi-classical analysis is displayed in Fig.~\ref{diff3}. It shows clearly
that the sequence of Vibrational levels approaches the semi-classical limit 1.11292. The departure from this
 limit at the uppermost side of the spectrum is stronger for the $1_u$ than for the $0_g^-$ 
state. A crossover region is spanned by the semi-classical limit and appears to be narrower for the  $1_u$
than for the $0_g^-$ case. 

\subsection{Test with Lennard-Jones molecules}
We apply our methodology to the Asymmetric Lennard-Jones case (ALJ) as considered 
by Trost {\it et al.} ~\cite{Trost98}.
After obtaining good agreement between the CFM eigenvalues and Trost {\it et al.}'s 
we display the results of the semi-classical analysis in Fig.~\ref{fig3}. 
 The sequence of neighbouring levels approaches the theoretical limit 1.92763. The departure
 from this limit at the uppermost side of the spectrum is stronger for the Trost case than 
for the $(C_1=1, C_2=3000)$ ALJ case and the width of 
the crossover region is almost zero in the Trost {\it et al.} case.

Gao ~\cite{Gao99} applied this analysis to sequences of neighbouring levels with his own 
calculated levels and claimed that BCP breaks down for all quantum systems in which the asymptotic 
interaction is of the form $1/r^n$ with $n > 2$.\\

Additionally Gao found that the agreement between his calculated levels and the SCF is better around
the middle of the vibrational spectra contrary to Stwalley {\it et al.}'s ~\cite{Stwalley78} and our work.

Fig. 2 clearly shows that the semi-classical limit is reached as we increase the quantum number for the
$0^-_g$ and $1_u$ states of the $^{23}{\rm Na}_2$ molecule. The departure is obtained with
the levels we obtained at the high end of the spectra with the VSCA method based on the CFM. The
crossover region is wider in the $0^-_g$ case and the departure is smaller.

On the other hand, Fig. 3 pertaining to the Lennard-Jones ($V(r) \sim -1/r^6$) case shows that
the crossover region is quite narrow for the ALJ as well as the Trost {\it et al.} case in comparison
with the  $^{23}{\rm Na}_2$ case. The departure is also stronger in the Trost case.
Additional work is needed is order to tie the width of the crossover region
to the asymptotic character of the potential.\\

\begin{figure}[htbp]
\begin{center}
\scalebox{0.5}{\includegraphics{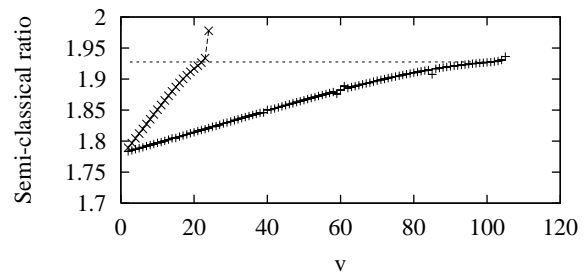}}
\end{center}
\caption{Approach of the semi-classical limit for the ALJ $(C_1=1, C_2=3000)$ potential 
as well as the Trost {\it et al.}  ~\cite{Trost98} model (upper curve) with the  vibrational quantum number.}
\label{diff6}
\end{figure}

\section{Exchange: local and non-local}

Calculations of cross sections for photoionisation require the resolution of
the integro-differential equations satisfied by the radial parts of the free
electron wavefunctions. Very often a simplifying approximation is made
through replacement of the exchange operator by a central equivalent exchange
potential (Furness and McCarthy \cite{Furness}, Bransden and Noble \cite{Bransden}). 
While this is probably satisfactory for electrons with energies of a few eV and
more it may cause problems near to threshold. Furthermore, to interpret the
new generation of experiments using polarized electrons the proper inclusion
of exchange may well be important. There exists a means of exact solution of
the equation for the radial wavefunction including static and static-exchange
terms (the diagonal parts of the direct potential and the exchange operator)
and a polarisation potential. This is the Distorted Wave Polarised Orbital
(DWPO) method of McDowell \textit{et al} \cite{McDowella} which was developed for Distorted
Wave Born calculations of excitation of Hydrogen (McDowell \textit{et al} \cite{McDowellb}) and
of Helium \cite{Scott} and replaces the integro-differential
equation by coupled differential equations. In both cases, only an s-state of
the atom is considered and the central potentials are expressed in analytical
form.

The phaseshift and the wavefunction in the asymptotic region were determined
by use of the analytic second order JWKB solution (Burgess \cite{Burgess}). 
The solutions are started by means of series expansions at the origin and continued by 
Numerov \cite{Numerov} integration: the form of series corresponding to the regular solution is
imposed by taking a power $r^{l'+1}$ at the origin in the $l'$
partial wave. For small $k$ the choice of integration step and
changeover point is delicate, since convergence of the series
requires small r whereas the Numerov integration becomes unstable (picking up
some of the irregular solution) if it is started at too small a radius.

Accuracy in evaluating the scattering phaseshift produced in electron atom collisions
is demanded in order to estimate reliably the cross section \cite{Raw}. 
In (e, 2e) \cite{Whelan} 
collision  type experiments, triple Differential Cross
Sections (TDCS) for electron impact ionization of an atom or ion, estimated within 
the Distorted Wave Born Approximation (DWBA), are required.  \\
Computational difficulties such as instabilities, lack of accuracy or convergence slow-down
arise in solving the corresponding Schr\"{o}dinger equation. The latter has to respect Pauli exclusion 
principle stating that the wave function of the incident electron be anti-symmetric with the wave
functions of the electrons in the target atom or molecule. In the
Hartree-Fock formulation, this requirement leads to the presence of non local
terms in the resulting coupled differential equations satisfied by the radial parts of the free
electron wavefunctions \cite{Raw,Whelan}. The non-local kernel of the integro-differential 
type equation originating from the exchange operator is often simplified and 
a central LEE potential is used or some decoupling 
procedure is employed \cite{Henry}. A review of the different methods to treat 
this problem on the basis of the Lippman-Schwinger equation is given 
by Rawitscher \textit{et al.} \cite{Raw}. 
Furness and McCarthy \cite{Furness} and  Bransden and Noble \cite{Bransden} cover the 
DWBA aspects. \\

Using a local exchange potential approximation is probably satisfactory for 
electrons with energies of a few eV and
more, however it may cause problems near to threshold. On the other hand,
the interpretation of experiments using polarized electrons, requires  
proper account for exchange. In addition, it is worthwhile to 
develop a version of the DWBA \cite{Whelan} including
non-local exchange and applicable to heavy atom targets.
This stems from the fact, the
successful Converged Close Coupling (CCC) approach of Bray \cite{Bray}
becomes increasingly difficult to apply as target complexity increases.\\

The Distorted Wave Polarised Orbital (DWPO) method is a means
for extracting the exact solution of
the radial wave equation with static, static-exchange
terms (the diagonal parts of the direct potential and the exchange operator)
and some polarisation potential. It has been developed by 
McDowell \textit{et al} \cite{McDowella} for DWBA
calculations for the ionization of Hydrogen (McDowell \textit{et al}
\cite{McDowellb}) and of Helium \cite{Scott}. It replaces the integro-differential
equation by coupled differential equations. Only an s-state of
the Hydrogen or Helium atoms is considered and the central potentials are 
expressed in analytical form. \\

The phaseshift and the wavefunction in the asymptotic region are determined
by use of the analytic second order JWKB solution (Burgess \cite{Burgess}). If the
energy of the free electron is $k^2$ Ry, the integro-differential
equation can be rewritten in terms of the variable $k r$ but the
value of the radius out to which exchange and the short-range part of the 
static potential remain significant, is determined by the extent of the electron 
cloud of the atom. For small $k$, the second order JWKB solution is valid only 
at radii very much larger than that of the asymptotic zone where 
exchange and short-range potentials are negligible. \\

\subsection{The integro-differential equation}

We consider the impact of a free electron of energy $k^2$ Ry on
a one-electron atomic system of nuclear charge $Z$ in a $1s$ atomic state of
energy $E_{10}$ Ry with a radial wavefunction $R_{10}(r) = 2 Z^{3/2} r \exp (- Z r)$. 
For a free electron with angular
momentum quantum numbers ${l' , m'}$, if we include only on-diagonal
potentials and exchange operators and replace the neglected off-diagonal
coupling potentials by a polarisation potential $V_{pol} ( r )$,
its radial wavefunction $F_{l'} ( k , r )$ satisfies the
integro-differential equation:

\begin{widetext}

\begin{equation}
\left[ {\frac{\partial^2}{\partial r^2}} -{\frac{l' ( l' + 1
)}{r^2}} + k^2 + V_{1 s} ( r ) + V_{pol} ( r )
+ W_{l'} ( r ) \right] F_{l'} ( k , r ) = 0
\end{equation}
 
where:

\begin{equation}
V_{1 s} ( r )  ={\frac{2 Z}{r}} - {\frac{2}{r}}{\int^r_0} \left| R_{10} ( r'
) \right|^{2} d r' - {\int^{\infty}_r}{\frac{2}{r'}} \left| R_{10} ( r' ) \right|^{2} d r' 
\end{equation}

and: 

\begin{eqnarray}
W_{l'} ( r ) F_{l'} ( k, r ) = ( - 1 )^{S + 1} R_{10} ( r )
\left\{ \left[ E_{10} - k^2 \right] \delta_{l' , 0} {\int^{\infty}_0} R_{10}
( r' ) F_{l'} ( k , r' ) d r'\right. \nonumber   \\
\left. - {\frac{2}{r}}{\int^r_0} R_{10} ( r' ) F_{l'} ( k , r' ) d r' -
{\int^{\infty}_r} {\frac{2}{r'}} R_{10} ( r' ) F_{l'} ( k , r' ) d r' \right\} 
\end{eqnarray}

This integro-differential equation can be transformed into the following system
of coupled differential equations, which is the starting point for McDowell
\textit{et al} \cite{McDowella, McDowellb}:
\begin{eqnarray*}
{\frac{\partial^2}{\partial r^2}} F_{l'} ( k , r ) - \left[ {\frac{l'(l'
+ 1)} {r^2}} -{ \cal V}_{1 s} ( r ) \right] F_{l'} ( k ,
r ) \\ + ( - 1 )^S R_{10} ( r ) \left[ {\frac{2}{r}} {\frac{1}{2 l' + 1}} \right]
G_{l'} ( k , r )  = ( - 1 )^{S + 1} R_{10} ( r ) \delta_{l' , 0} A ( k)
\end{eqnarray*}
\begin{equation}
{\frac{\partial^2}{\partial r^2}} G_{l'} ( k, r ) -{\frac{l'(
l' + 1 )} {r^2}} G_{l'} ( k , r ) + {\frac{2 l' + 1}{r}}
R_{10} ( r ) F_{l'} ( k , r ) = 0
\end{equation}

\end{widetext}

where
\begin{equation}
A ( k ) = \left[ k^2 - E_{10} \right] {\int^{\infty}_0} R_{10} ( r' )
F_{l'} ( k , r' ) d r' 
\end{equation}
is a term appearing for the special case of zero orbital momentum ($l'=0$). 
$\delta_{l' , 0}$ is the Kronecker delta and 
\begin{equation}
{\cal V}_{1 s} ( r )= k^2 + V_{1 s} ( r ) + V_{pol} ( r )
\end{equation}

Our aim is to solve this system subject to the boundary conditions 
\cite{McDowella, McDowellb} for the Hartree functions:

\begin{widetext}

\begin{eqnarray}
& F_{l'} ( k , r ) \mathop{\longrightarrow}\limits_{r \rightarrow 0} 0  \hspace{5.3cm}  
  &  G_{l'} ( k , r ) \mathop{\longrightarrow}\limits_{r \rightarrow 0} 0  \nonumber  \\
& F_{l'} ( k , r )  \mathop{\longrightarrow}\limits_{r \rightarrow \infty} a_{l'} ( k )
\left\{ s_{l'} ( k r ) - \tan \left[ \delta_{l'} ( k ) \right] c_{l'} ( k r) \right\}
&  G_{l'} ( k , r ) \mathop{\longrightarrow}\limits_{r \rightarrow \infty} 0 
\end{eqnarray}

\end{widetext}

where $a_{l'} ( k )$ is a normalisation factor, $\delta_{l'} ( k)$ 
is the phaseshift for specific $\{ k , l' \}$ and
$ s_{l'} ( \rho )$ and $c_{l'} ( \rho )$ are
respectively:
\begin{itemize}
\item
$\rho$ multiplied spherical Bessel and Neumann functions when $Z =
1$ (so that ${\cal V}_{1 s} ( r )$ is a short range potential
falling off faster than $r^{- 1}$ as $r$ tends to infinity);
\item
regular and irregular Coulomb wavefunctions when $Z > 1$
(so that ${\cal V}_{1 s} ( r )$ is a long range potential behaving
like $( Z - 1 ) r^{- 1}$ when $r$ tends to infinity).
\end{itemize}

In the more general case of an ion with a frozen core and an outer shell of
electrons in $\{n , l \}$ states of radial wavefunction
$R_{nl} ( r )$, and a free electron in the state
$\{ k , l' \}$ we have a larger set of coupled equations:

\begin{widetext} 

\begin{eqnarray}
{\frac{\partial^2}{\partial r^2}} F_{l'} ( k , r ) - \left[{\frac{l' (
l' + 1 )}{r^2}} -{\cal V}_{n l} ( r ) \right] F_{l'} (
k , r ) \nonumber \\ - ( - 1 )^S R_{nl} ( r ){\frac{2}{r} \sum_{\lambda}
J_{l , l' , \lambda}} G_{l'}^{\lambda} ( k , r )
= ( - 1 )^{S + 1} R_{n l} ( r ) \delta_{l , l'} A_{n l , l'} ( k ) \nonumber  \\
{\frac{\partial^2}{\partial r^2}} G_{l'}^{\lambda} ( k , r )
-{\frac{\lambda ( \lambda + 1 )}{r^2}} G_{l'}^{\lambda} ( k
, r ) + {\frac{2 \lambda + 1}{r}} R_{n l} ( r )  F_{l'} ( k , r )
= 0
\end{eqnarray}

where:

\begin{eqnarray}
{\cal V}_{n l} ( r ) =  k^2 - V_{nl} ( r ) & \nonumber   \\
- 2 \sum_{\lambda} I_{l , l' ,
\lambda} \left\{ {\int^r_0}{\frac{r'^{\lambda}}{r^{\lambda + 1}}} \left| R_{n
 l} ( r' ) \right|^{2} d r'
+{\int^{\infty}_r}{\frac{r^{\lambda}}{r'^{\lambda + 1}}} \left| R_{n l} ( r'
) \right|^{2} d r' \right\}   + V_{pol} ( r ) &
\end{eqnarray}
and:
\begin{equation}
A_{n l , l'} ( k ) =  \left[k^2 - E_{n l} \right] {\int^{\infty}_0}
R_{n l} ( r' ) F_{l'} ( k , r' ) d r' 
\end{equation}

These equations are subject to the boundary conditions, for all possible $\lambda$ values:
\begin{eqnarray}
& F_{l'} ( k , r ) \mathop{\longrightarrow}\limits_{r \rightarrow 0} 0  \hspace{5.3cm}
& G_{l'}^{\lambda} ( k , r ) \mathop{\longrightarrow}\limits_{r \rightarrow 0} 0  \nonumber
\\
&  F_{l'} ( k , r )  \mathop{\longrightarrow}\limits_{r \rightarrow \infty}
a_{l'} ( k ) \left\{ s_{l'} ( k r ) - \tan \left[ \delta_{l'} ( k ) \right] c_{l'}
( k r ) \right\}  
&  G_{l'}^{\lambda} (k , r ) \mathop{\longrightarrow}\limits_{r \rightarrow \infty} 0
\label{eq4}
\end{eqnarray}

\end{widetext}

$I_{l , l' , \lambda}$ and $ J_{l , l' , \lambda}$ are angular integrals which 
depend on the number of electrons in the ion outer shell and the angular 
momentum coupling scheme. $V_{nl} ( r )$ is a central potential for attraction of an
electron by the core and $E_{n l}$ is the total energy of the outer
shell electrons. This is not applicable to hydrogenic ions as the degeneracy of the
energy in $l$ makes it essential to include channel coupling potentials.
We will not consider it further here except to note that $ J_{l , l' ,\lambda}$ 
imposes the triangular rule $\left| l - l' \right| \leq \lambda \leq l + l'$ and
 $l + l' + \lambda$ even. This gives an idea of 
the number of different $G_{l'}^{\lambda}$ present and the  extent of the 
problem we ultimately wish to solve, in the case of more complex atoms.

\subsection{The CFM for solving the DWPO equations}

In order to facilitate the presentation it is convenient to use $f_1$ in place of 
$F_{l'}$ and $f_2$ in place of $G_{l'}$ so as to rewrite the coupled equation 
system (2) as a special case of the more general system:
\begin{eqnarray}
f_1 '' ( r ) + V_{11} ( r ) f_1 ( r ) + V_{12} ( r ) f_2 ( r ) & = 
\delta_{l', 0} A ( k ) W_1 ( r )  \nonumber \\
f_2 '' ( r ) + V_{22} ( r ) f_2 ( r ) + V_{21} ( r ) f_1 ( r ) & = 
\delta_{l', 0} A ( k ) W_2 ( r )  \nonumber \\
\end{eqnarray}
with

\begin{widetext}
\begin{eqnarray}
V_{11} ( r ) = {\cal V}_{1 s} ( r ) - \displaystyle{\frac{l' ( l' + 1)}{r^2}},
& & 
V_{12} ( r ) = ( - 1 )^S R_{10} ( r ) \left[ \displaystyle{\frac{2}{r}} {\frac{1}
{(2 l' + 1)}} \right],
    \nonumber \\
V_{21} ( r ) = \displaystyle{\frac{(2 l' + 1)}{r}} R_{10} ( r ), 
& &  
V_{22} ( r ) = - \displaystyle{\frac{l' ( l' + 1 )}{r^2}}, 
 \nonumber  \\
W_1 ( r ) = ( - 1 )^{S + 1} R_{10} ( r ),
& &  
W_2 ( r ) = 0
\end{eqnarray}
We can construct the general solution of the above equations as  
(see Kobeissi and Fakhreddine~\cite{Kobeissi91a}):
\begin{eqnarray} 
f_1 ( r ) & = f_1 ( r_0 ) \alpha_{11} ( r ) + f_1 ' ( r_0 ) \beta_{11} ( r )
 + f_2 ( r_0 ) \alpha_{12} ( r ) + f_2 ' ( r_0 ) \beta_{12} ( r ) + \delta
_{l' , 0} A ( k ) \sigma_1 ( r )  \nonumber \\
f_2 ( r ) & = f_1 ( r_0 ) \alpha_{21} ( r ) + f_1 ' ( r_0 ) \beta_{21} ( r )
 + f_2 ( r_0 ) \alpha_{22} ( r ) + f_2 ' ( r_0 ) \beta_{22} ( r ) + \delta
_{l' , 0} A ( k ) \sigma_2 ( r )
\end{eqnarray}

\end{widetext}

where $\left\{ \alpha_{1 j} ( r ) , \alpha_{2 j} ( r ) \right\}$ and
$\left\{ \beta_{1 j} ( r ) , \beta_{2 j} ( r ) \right\}$ are two
different pairs of independent solutions of the homogeneous system:
\begin{eqnarray}
g_1 '' r ) + V_{11} ( r ) g_1 ( r ) + V_{12} ( r ) g_2 ( r )  & = 0   \\
g_2 '' ( r ) + V_{22} ( r ) g_2 ( r ) + V_{21} ( r ) g_1 ( r ) & = 0
\label{eq11}
\end{eqnarray}
satisfying the initial conditions at an arbitrary point $r = r_0$
\begin{eqnarray}
\alpha_{i j} ( r_0 ) = \beta_{i j} ' ( r_0 ) = \delta_{i , j},
& &
\alpha_{i j} ' ( r_0 ) = \beta_{i j} ( r_0 ) = 0   \nonumber
\label{eq12}
\end{eqnarray}
and $\{ \sigma_1 ( r ) , \sigma_2 ( r ) \}$ is a particular solution
of the inhomogeneous system:
\begin{eqnarray}
h_1 '' ( r ) + V_{11} ( r ) h_1 ( r ) + V_{12} ( r ) h_2 ( r ) & = W_1 ( r)  \\
h_2 '' ( r ) + V_{22} ( r ) h_2 ( r ) + V_{21} ( r ) h_1 ( r ) & = W_2 ( r )
\label{eq13}
\end{eqnarray}

satisfying the initial conditions at $r = r_0$ 
\begin{equation}
\sigma_i ( r_0 ) = \sigma_i' ( r_0 ) = 0
\label{eq14}
\end{equation}

The general solution in matrix form is written as: 
\begin{equation}
Y ( r ) = \alpha ( r ) Y ( r_0 ) + \beta ( r ) Y ' ( r_0 ) + \delta_{l' , 0}
A ( k ) \sigma ( r )
\label{eq15}
\end{equation}
with

\begin{widetext}

\begin{eqnarray}
Y ( r ) = \left( \begin{array}{c} f_1 ( r )\\ f_2 ( r )\\ \end{array} \right),
&  
\alpha ( r ) = \left( \begin{array}{cc} \alpha_{11} ( r ) & \alpha_{12} 
( r )\\ \alpha_{21} ( r ) & \alpha_{22} ( r )\\ \end{array} \right),
& 
\beta ( r ) = \left( \begin{array}{cc} \beta_{11} ( r ) & \beta_{12}
( r )\\ \beta_{21} ( r ) & \beta_{22} ( r )\\ \end{array} \right)
\end{eqnarray}

\end{widetext}

Each column of the (2x2) matrices $\alpha(r)$ and $\beta(r)$ is a particular solution
of \ref{eq11} with initial values given by \ref{eq12}. The column matrix:$\sigma ( r ) = 
\pmatrix{\sigma_1 ( r ), \sigma_2 (r )} $ is a particular solution of \ref{eq13}
with initial condition given by \ref{eq14}.
Using the boundary conditions \ref{eq4} imposes:
$\alpha ( 0 ) Y ( r_0 ) + \beta ( 0 ) Y ' ( r_0 ) + \delta_{l' , 0} A ( k) \sigma ( 0 ) = 0$,
which leads to 
$\beta^{- 1} ( 0 ) \alpha ( 0 ) Y ( r_0 ) + Y ' ( r_0) + \delta_{l , 0} A ( k ) \beta^{- 1} ( 0 ) \sigma ( 0 ) = 0 $ 
where
$\beta^{- 1} ( r )$ is the inverse of the matrix $\beta ( r )$.

Thus, the constant matrices at point $r=r_0$, $Y ' ( r_0 ) $ and $Y ( r_0 )$ are related by:
\begin{equation}
Y ' ( r_0 ) = Y ( r_0 ) \Lambda + \delta_{l' , 0} A ( k) \lambda,
\end{equation}

where the matrices $\Lambda$ and $\lambda$ are given by:

\begin{eqnarray}
\Lambda = - \beta^{- 1} ( 0 ) \alpha ( 0 ),
& 
\lambda = - \beta^{-1} ( 0 ) \sigma ( 0 )
\end{eqnarray}

Substituting back into \ref{eq15} we then get:
\begin{equation}
Y ( r ) = \varphi ( r ) Y ( r_0 ) + \delta_{l' , 0} A ( k ) \gamma ( r)
\end{equation}

the functions $\varphi(r)$ and $\gamma(r)$ are related to the particular solutions
$\alpha(r), \beta(r)$ and $\sigma(r)$ by:
\begin{eqnarray}
\varphi ( r ) = \alpha ( r ) + \beta ( r ) \Lambda,
& 
\gamma ( r ) =\beta ( r ) \lambda + \sigma ( r )
\end{eqnarray}

We notice again that $\varphi(r)= \pmatrix {\phi_1( r ) , \phi_2( r )} $ 
and $\gamma(r)=\pmatrix {\gamma_1( r ) , \gamma_2 ( r )}$
are particular solutions of  system (8) since they are linear combinations of $\alpha(r)$ and
$\beta(r)$.
 
The initial values at the arbitrarily chosen starting point $r =r_0$ are:
\begin{equation}
\varphi ( r_0 ) = I , \varphi ' ( r_0 ) = \Lambda,  \gamma ( r_0 )= 0, \gamma ' ( r_0 ) = \lambda  
\end{equation} 

where $I$ is the unit matrix. The solution constructed from $\varphi ( r )$ 
and $\gamma ( r )$ is a particular 
solution of the coupled equations (8) for which the functions 
$\left\{ f_1 (r ) , f_2 ( r ) \right\}$ are regular at the origin.

\subsection{Phaseshift calculation}

From the first of boundary conditions we can determine the phaseshift $\delta_l$ from:
\begin{eqnarray}
\tan \delta_{l'} = \lim_{r \rightarrow \infty} Q ( r ), \mbox{   where: }  \nonumber \\
Q ( r ) =  - \displaystyle{\frac{f_1 ' ( r ) s_{l'} (k r ) - f_1 ( r ) k s_{l'} ' ( k r )}
{f_1 ' ( r ) c_{l'} ( k r ) - f_1 ( r ) k c_{l'} ' ( k r )}}
\label{eq26}
\end{eqnarray}

We follow Kobeissi \textit{et al} \cite{Kobeissi91} in using the recursion relations for
$s_{l'} ( \rho )$ and $c_{l'} ( \rho )$ :

\begin{widetext}
\begin{eqnarray}
Z=1 : \nonumber \\
 Q ( r ) = \frac{\left[ f_1 ' ( r ) -{\frac{( l' + 1 )}{r}} f_1 ( r )
\right] s_{l'} ( k r ) + k f_1 ( r ) s_{l' + 1} ( k r )}{\left[ f_1 '
( r ) -{\frac{( l' + 1 )}{r}} f_1 ( r ) \right] c_{l'} ( k r ) + k f_1
( r ) c_{l' + 1} ( k r )}   \nonumber \\
Z>1 :   \nonumber \\
Q ( r ) =  \frac{\left[ f_1 ' ( r ) + \left\{ {\frac{( Z - 1 )}{k ( l' +
1 )}} -{\frac{( l' + 1 )}{r}} \right\} f_1 ( r ) \right] s_{l'} ( k r ) +
\sqrt{k^2 +{\frac{( Z - 1 )^2}{( l' + 1 )^2}}} f_1 ( r ) s_{l' + 1} ( k
r )}{\left[ f_1 ' ( r ) + \left\{ {\frac{( Z - 1 )}{k ( l' + 1 )}} -{\frac{(
l' + 1 )}{r}} \right\}  f_1 ( r ) \right] c_{l'} ( k r ) + \sqrt{k^2
+{\frac{( Z - 1 )^2}{( l' + 1 )^2}}} f_1 ( r ) c_{l' + 1} ( k r )}
\label{eq27} 
\end{eqnarray}

\end{widetext}

\vspace{1cm}

The function $Q(r)$ can be obtained for any radius $r$. We calculate it
at large $r$ values and examine its behaviour. When it tends to a
constant limit, we consider that the asymptotic region has been reached and
the phaseshift is determined modulo $2 \pi $. 

Numerically, the phaseshift is actually calculated with the following steps:

\begin{enumerate}

\item An arbitrary point $r_0$ is chosen as the origin to proceed with the integration.

\item  The canonical functions $\alpha(r)$, $\beta(r)$ and $\sigma(r)$ are computed for 
$r< r_0$ : Eqs \ref{eq11} and \ref{eq13} are integrated starting at $r_0$ with 
$\alpha(r)$, $\beta(r)$ and $\sigma(r)$ initialised with \ref{eq12} and \ref{eq14}. The matrices $\Lambda$
and $\lambda$ are deduced. Progressing toward the origin $r=0$, the integration is stopped when
the matrices $\Lambda$ and $\lambda$ attain stable values. Numerically, this is done by monitoring
the values of their determinants.

\item The computation of $\Lambda$ and $\lambda$ serves to calculate the other canonical functions
 $\varphi(r)$ and $\gamma(r)$ and consequently estimate $A_1$ and $A_2$. 
 
 \item Repeating the computation with $r \ge r_0$ allows to calculate the ancillary functions
 $f_1(r)$, $f'_1(r)$ and $Q(r)$ (relations \ref{eq26} and \ref{eq27}). The calculation is stopped when
 $Q(r)$ attains a stable value corresponding to $\tan\delta$. The stability of the phaseshift value
is attained when the ratio $|Q(r_{p+1})-Q(r_p)|/|Q(r_{p+1})+Q(r_p)| \le \epsilon_S$, where 
$[r_p, r_{p+1}]$ is the interval obtained after $p$ integration steps 
and $\epsilon_S$ is termed the stability error.

 \end{enumerate}

 The phaseshift overall error (desired error or tolerance)
depends on both  $\epsilon_T$ and  $\epsilon_S$ as well as computer/compiler arithmetic,
roundoff errors and intermediate algorithmic/numerical operations. We estimate it through
a comparison with analytical or other available results 
(such as asymptotic or at special points). \\

For small $k$ the outward Numerov integration of the regular solution can
get out of control if it picks up a tiny fraction of the irregular
solution, because of ill-conditioning due to round-off errors. An alternative
is to modify the potential by introducing a hard core: ${\cal V}_{n l} (r )$ 
is set artificially to infinity for $r < r_s$, where
$r_s$ is the starting point of the integration and retains its
original form for $r > r_s$. \\
As mentioned by Bayliss \cite{Bayliss} \textit{et al}, this
method gives results sensitive to the point in the classically forbidden
region at which the integration is started. If the starting point is too
small some solutions become unstable; if it is too large for the initial
conditions used, the solution is inaccurate. 
In the CFM, the solutions $\alpha ( r )$, $\beta ( r)$ and $\sigma ( r )$ 
initially generated are in fact linear
combinations of the regular and irregular solutions of the coupled equations
and by taking the linear combinations $\varphi ( r )$ and $\gamma ( r)$ 
we eliminate from them the irregular solution.

\subsection{Phaseshift accuracy and comparison to the S-IEM}

We apply the present method to the case of the collision of a low-energy electron
with atomic hydrogen, \textit{i.e.} generating the wavefunctions needed for DWBA
calculations of electron impact ionization of atomic hydrogen. In this case
the static potential is given by:
\begin{equation}
V_{1 s} ( r ) = - 2 \left( 1 + \frac{1}{r} \right) \exp ( - 2 r )
\end{equation}
Like McDowell and collaborators, we use a Callaway-Temkin polarisation
potential (see Drachman and Temkin \cite{Drachman}) of the form:
\begin{widetext}
\begin{equation}
V_{pol} ( r ) = - {\frac{9}{2 r^4}} \left[ 1 - e^{- 2 r} \left( 1 + 2
r + 2 r^2 + \frac{4}{3} r^3 + \frac{2}{3} r^4 + \frac{4}{27} r^5 \right) 
\right]
\end{equation}
\end{widetext}

To integrate the coupled equation system, preference is given in
the present work to the VSCA method described above. The latter method  was
shown by Kobeissi \textit{et al} \cite{Kobeissi88, Kobeissi91b, Fakhreddine99}   
to be highly accurate in the case of both single and
coupled differential equations. It requires potentials to be expressed in
analytical form. Numerical potentials can generally be fitted by piecewise
analytical or polynomial functions such as cubic splines. The integration can
be safely taken out to a very large radius, where ${\cal V}_{1 s} ( r)$
assumes its asymptotic form and the phaseshift determined.
We refer to this method as the CFM Exact Exchange (CFMEE). 

In order to further gauge the reliability of our results, we performed a series 
of calculations varying the stability error $\epsilon_S$ and the truncation error $\epsilon_T$
in order to test the accuracy and
convergence speed of our method. The tests are displayed in the tables below
for the singlet case:

\begin{widetext}

\begin{table}[h!]
\centering
\caption{Singlet ($S=0, l'=0$) phaseshifts for e-H  scattering versus momentum as the 
stability error is decreased while truncation error $\epsilon_T$ is fixed at 10$^{-14}$ (upper sub-table) 
and as the truncation error is decreased with a fixed stability error $\epsilon_S=10^{-14}$ (lower sub-table)}

\begin{tabular}{c | c c c c c c}
\hline
   &   \multicolumn{6}{c}{Stability error $\epsilon_S$} \\
  $ k$       & 10$^{-4}$  &  10$^{-6}$ & 10$^{-8}$ & 10$^{-10}$  & 10$^{-12}$  & 10$^{-14}$ \\
\hline
0.1 & 2.54421122 & 2.53027301 & 2.5274412 & 2.52744125 & 2.52744125 & 2.52744125 \\
0.2 & 2.06575195 & 2.03935101 & 2.03407088 & 2.03407088 & 2.03407088 & 2.03407088 \\
0.3 & 1.70757716 & 1.67206099 & 1.66518977 & 1.66518968 & 1.66518968 & 1.66518968 \\
0.4 & 1.42957123 & 1.3914665 & 1.38497592 & 1.38497592 & 1.38497592 & 1.38497592 \\
0.5 & 1.2095031 & 1.17365026 & 1.16825708 & 1.16825708 & 1.16825708 & 1.16825708 \\
0.6 & 1.03735556 & 1.00552357 & 1.00072357 & 1.00072404 & 1.00072404 & 1.00072404 \\
0.7 & 0.90616322 & 0.87811171 & 0.87375939 & 0.87376024 & 0.87376024 & 0.87376024 \\
0.8 & 0.80734542 & 0.78365856 & 0.77961723 & 0.77961723 & 0.77961723 & 0.77961723 \\
0.9 & 0.75274001 & 0.75420532 & 0.71342049 & 0.71342026 & 0.71342027 & 0.71342027 \\
1.0 & 0.69239713 & 0.7010382 & 0.6701203 & 0.6701203 & 0.6701203 & 0.6701203 \\
\hline
   &   \multicolumn{6}{c}{Truncation error $\epsilon_T$} \\
  $ k$       & 10$^{-4}$  &  10$^{-6}$ & 10$^{-8}$ & 10$^{-10}$  & 10$^{-12}$  & 10$^{-14}$ \\
\hline
0.1 & 3.00127274 & 3.0307537 & 2.52412523 & 2.52463171 & 2.52558189 & 2.5274412   \\
0.2 & 2.82900282 & 2.88186498 & 2.02681296 & 2.02795407 & 2.03007799 & 2.03407088  \\
0.3 & 2.65292424 & 1.65536145 & 1.65611743 & 1.65755527 & 1.66025047 & 1.66518977  \\
0.4 & 2.4734271 & 1.37484701 & 1.37567631 & 1.37717447 & 1.37993599 & 1.38497592  \\
0.5 & 1.15828993 & 1.15870411 & 1.15948866 & 1.16092162 & 1.16354419 & 1.16825708  \\
0.6 & 2.14501974 & 0.9920774 & 0.99279684 & 0.9941118 & 0.99649203 & 1.00072357  \\
0.7 & 0.86558783 & 0.8658675 & 0.86653892 & 0.86774235 & 0.86992432 & 0.87375939  \\
0.8 & 1.8745865 & 0.77331792 & 0.7738408 & 0.77481281 & 0.77655005 & 0.77961723  \\
0.9 & 1.72900321 & 0.70870253 & 0.70910062 & 0.70982718 & 0.71113237 & 0.71342049  \\
1.0 & 1.60803611 & 0.66656396 & 0.66686498 & 0.66741118 & 0.6684008 & 0.6701203  \\
\hline
\end{tabular}
\label{tol}
\end{table}

\end{widetext}

Recently, Rawitscher \textit{ et al.} \cite{Raw} obtained very accurate results for the phaseshift at
$l'=0$ for momentum $k=0.2/a_0$ in the absence of any polarization potential
but with rigorous inclusion of the Fock exchange term. They obtained the phaseshifts 
and scattering lengths in the singlet and triplet states. The method they use,
the S-IEM (Spectral Integral Equation Method), is based on partitioning the
integration interval $[0, r_{max}]$ into a fixed number of partitions expressing
the integration kernel with a set of four different functions. This  transforms the problem into
a block tridiagonal matrix system built from sets of 4x4 block matrices corresponding
each to a given partition \cite{Raw}. \\  
The phaseshift expressed in the form $k/\tan(\delta)$ versus momentum $k$ is displayed  
for the singlet and the triplet in Fig. 5. Our results (crosses) fall exactly on top
of the continuous curves for the singlet and the triplet obtained by
Rawitscher \textit{ et al.} \cite{Raw} with the S-IEM. Whenever, either of the curves cross the horizontal
axis, the phaseshift value is $\pi/2$ modulo ($\pi$) and  $\tan(\delta)$ is singular.
The CFM method like the S-IEM, had no particular difficulty (instability or
slowing down) in reproducing the singularity
and the phaseshifts were obtained as described in section 5.1, for arbitrarily 
small values of the momentum $k$ ($k \ge 10^{-8}$ in $1/a_0$ units).

\begin{widetext}

\begin{figure}[h!]
\centering
\scalebox{0.5}{\includegraphics[angle=0]{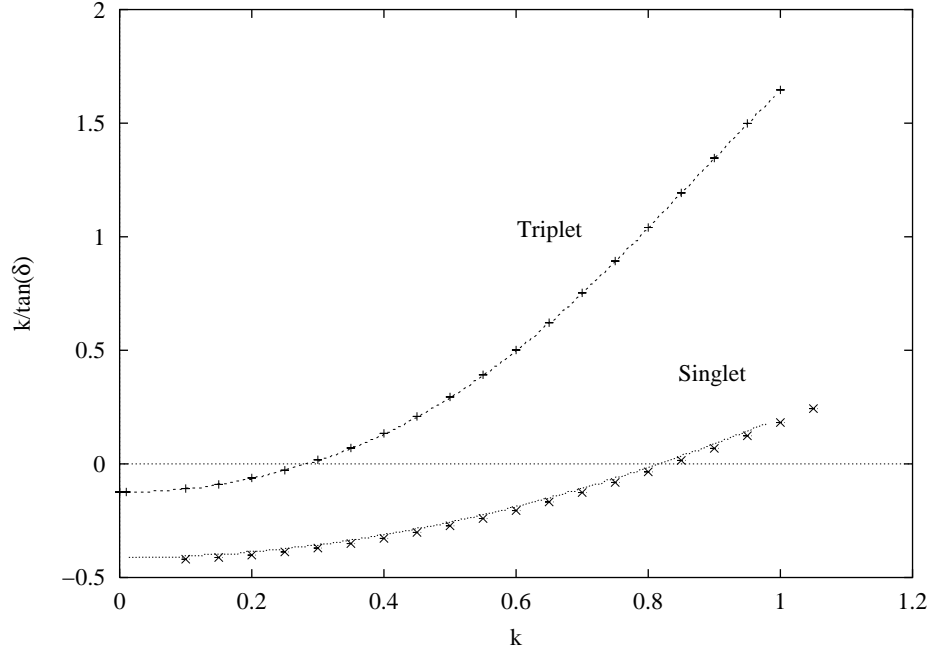}}
\caption{Singlet and Triplet $l'=0$, CFMEE results (crosses) values of $k/\tan(\delta)$ 
versus incident momentum $k$ in $1/a_0$ units and comparison to  Rawitscher \textit{et al.} 
(continuous lines). The smallest value used for $k$ is 10$^{-8}$.}
\label{dwp5}
\end{figure}

\end{widetext}

The rate of convergence as a function of stability error  $\epsilon_S$ of the CFM result for the triplet
phaseshift at $k=0.2/a_0$ is displayed in table \ref{trip02} along with the 
S-IEM result. It is remarkable to observe that as the stability error is decreased,
we obtain steady convergence toward the S-IEM result. If we consider, in that case, the S-IEM result
as exact, we can infer that the overall desired error (tolerance) behaves as 
$\sim {\epsilon_S}^{0.45}$, implying that  $\epsilon_S$ is a good indicator of 
accuracy despite its deterioration across intermediate arithmetic operations.

\begin{table}[h!]
\centering
\caption{Triplet phaseshift accuracy as the stability error is decreased at $k=0.2/a_0$. 
S-IEM is Rawitscher \textit{et al.} Spectral Integral Equation Method}
\begin{tabular}{c c } 
\hline
 Stability error $\epsilon_S$  &  $\delta$ \\
\hline
10$^{-4}$   &   1.8927821 \\
10$^{-5}$    &  1.8773093 \\
10$^{-6}$   &  1.8724465 \\
10$^{-7}$   &    1.8708870 \\
10$^{-8}$   &   1.8703880 \\
10$^{-9}$   &  1.8702301  \\
10$^{-12}$   &  1.8701771  \\
S-IEM       &    1.8701579 \\
\hline
\end{tabular}
\label{trip02}
\end{table}

The scattering length $a$ for the electron-hydrogen atom scattering is based on the low
momentum expansion of the scattering phaseshift:

\begin{equation}
k \cot \delta = -1/a + r_e k^2 + O(k^3)
\end{equation}

where $r_e$ the effective range \cite{Raw}. 
The left-hand side of the above expression is calculated for two very small values of the
momentum $k$ differing by a factor of two and the values of $a$ and $r_e$ are solved for \cite{Raw}. 
This procedure is repeated for decreasing values of $k$ and  
stability error $\epsilon_S$ until the numerical value of the result is obtained to a given
number of significant figures. In table \ref{singtol}, we display the results of the
CFM and S-IEM for the singlet case. The agreement between the S-IEM result for the scattering length
8.100312397 and the corresponding CFM result (8.100307932) is excellent. 

\begin{table}[h!]
\centering
\caption{Singlet scattering length versus momentum as the stability error is decreased. 
S-IEM is Rawitscher \textit{et al.} Spectral Integral Equation Method}
\begin{tabular}{c| c c c} 
\hline
    &   \multicolumn{3}{c}{Stability error $\epsilon_S$}   \\
\cline{2-4} 
 $ k$  &        10$^{-4}$  &  10$^{-6}$ & 10$^{-8}$ \\
\hline
 10$^{-2}$      & 8.058030140 &  8.104593677  & 8.109676748  \\
10$^{-3}$    &   8.081922310 &  8.098359718 &  8.100205405 \\
10$^{-4}$   &   8.093862531 &  8.099570349 &  8.100235983 \\
10$^{-5}$    &  8.098414768 &  8.100045256  & 8.100281954 \\
10$^{-6}$   &   8.100073345  & 8.100226775  & 8.100299303 \\
10$^{-8}$   &   8.100305162  & 8.100305163  &  8.100307932 \\
S-IEM    & & &  8.100312397 \\
\hline
\end{tabular}
\label{singtol}
\end{table}

The CFM result is obtained for a value of momentum 
$k= 10^{-8}/a_0$ and stability error $\epsilon_S=10^{-8}$, whereas  the S-IEM result was obtained with
a value  of momentum  $k= 10^{-5}/a_0$, and maximum integration range $r_{max} \sim 50$ with 
approximately 1000 mesh points. \\
Finally we display in table \ref{tripk0} the result for the triplet scattering length as momentum
$k$ decreases steadily toward $k= 10^{-8}/a_0$.

\begin{table}[h!]
\centering
\caption{Triplet scattering length (with 10$^{-8}$ stability error) as momentum decreases and comparison
 to the S-IEM result. Note the faster convergence than the singlet case.}
\begin{tabular}{c c } 
\hline
  k   &   a  \\
\hline
10$^{-2}$   &  2.3497331  \\
10$^{-3}$   &   2.3493995 \\
10$^{-4}$   &   2.3493961 \\
10$^{-5}$   &   2.3493961  \\
10$^{-6}$   &  2.3493961 \\
10$^{-7}$   &  2.3493961 \\
10$^{-8}$   &  2.3493961 \\
S-IEM       &  2.3493961 \\
\hline
\end{tabular}
\label{tripk0}
\end{table}

Roche \textit{et al.} \cite{Roche}
developed, recently,  an exact exchange method, within the DWBA, for the evaluation
of high and low energy  electron impact ionisation experiments on Hydrogen and Lithium,
with presence of polarisation. 
In the table below, we compare our results (CFMEE) to theirs
for the singlet and triplet s-wave phaseshifts at high energies. 
The polarisation used is of the form: $ V_{pol}(r)= - \frac{\alpha}{2 r^{4}} $.

\begin{table}[h!]
\begin{center}
\caption{Singlet s-wave phaseshifts with polarization $ V_{pol}(r)= - \frac{\alpha}{2 r^{4}} $
 and comparison to the CFMEE method.}
\begin{tabular}{c c c} 
\hline
 Energy in eV        &     CFMEE        &        Roche \textit{et al.} \cite{Roche} \\
\hline
    14.6      &          -1.46166         &      -1.46084 \\
    15.0      &          -1.4720848      &       -1.47127  \\
    15.6      &          -1.48717379     &        -1.48642 \\
    17.6      &          -1.53373849      &      -1.53297 \\
    20.0       &         - 1.58264557     &      -1.51896 \\
    25.0      &          -1.66616627      &      -1.66483 \\
   30.0      &           -1.73100847      &       -1.73061 \\
\hline
\end{tabular}
\end{center}
\label{singlet}
\end{table}

\begin{table}[h!]
\begin{center}
\caption{Triplet s-wave phaseshifts with polarization $ V_{pol}(r)= - \frac{\alpha}{2 r^{4}} $ 
 and comparison to the CFMEE method. }
\begin{tabular}{c c c} 
\hline
 Energy in eV        &     CFMEE        &        Roche \textit{et al.} \cite{Roche}  \\
\hline
    14.6      &           -1.45993029     &      - 1.43966   \\ 
    15.0      &           -1.46697595     &      - 1.44676  \\
    15.6      &           -1.47691298     &      - 1.45683 \\
    17.6      &           -1.50572363     &      - 1.48681 \\
    20.0      &           -1.53328328     &      - 1.51529 \\
    25.0     &            -1.57639125      &     - 1.56060 \\
    30.0      &           -1.60975092      &     - 1.59548  \\
\hline
\end{tabular}
\end{center}
\label{triplet}
\end{table}

The remarkable agreement shows that our method performs well
at high energies and presence of polarisation,
with methods  using exact exchange instead of local exchange.
Our work has also shown that a distorted wave code, with exact treatment of exchange and
Numerov integration, breaks down with appearance of numerical instabilities at
energies below about 5 eV. 
Codes with LEE potentiels give poor results for low partial waves at
energies below 15 eV.

\section{One-dimensional potential problems}

In the last section before the conclusion, we apply the CFM to a couple of
1D potentials to illustrate once again its capabilities beyond the focus of this
work mainly concerned with the 3D case.

The archetypical 1D problem is the infinite square well potential of width $w$ defined by: \\
$V(x)=0, \hspace{2mm}  0 < x < w , V(0)= \infty,  V(w)= \infty$, \\
the eigenvalues are given by: $E_n=\frac{\hbar^2}{2m_p}{(\frac{n\pi}{w})}^2, n=1,2...$
with $m_p$ is the particle mass (not necessarily an electron as  generally assumed). \\
In order to conform with the CFM general framework, one transforms the closed finite interval 
$[0,w]$ to the open infinite interval $[0,\infty[$. \\
It is interesting to note that while, in general, the CFM maps an arbitrary potential 
(infinite interval) over the infinite square well potential (finite interval), 
this problem might be viewed exactly as the reverse mapping. \\ 
With the use of the transformation
$r=\frac{x}{(w-x)}$  the  Schr\"odinger equation is transformed into:
\begin{equation}
-\frac{\hbar^2}{2m_p}\frac{d^2 \psi(x)}{d x^2}=E \psi(x), \hspace{2mm} 0 < x < w
\end{equation} 
with the boundary conditions: $ \psi(0)= 0,  \psi(w)= 0$.\\
The Schr\"odinger equation defined over the interval  $0 < r < \infty$
acquires a first-order derivative term becoming:
\begin{equation}
-\frac{\hbar^2}{2m_p}\left[\frac{{(1+r)}^4}{w^4}\frac{d^2 \psi(r)}{d r^2}
+ \frac{2{(1+r)}^3}{w^2}\frac{d \psi(r)}{d r}\right]
=E \psi(r),
\end{equation} 
The VSCA algorithm derived earlier for the standard Schr\"odinger equation must be altered into
(using units such that $\hbar=1, m_p=1$):
\begin{widetext}
\begin{equation}
 (n + 2) (n + 1) C_{n + 2}^{(p)} =  -2 \sum_{m = 0}^{n} (m+1) a_{n - m}^{(p)} C_{m+1}^{(p)}
 - 2E w^2 \sum_{m = 0}^{n} b_{n - m}^{(p)} C_{m}^{(p)}
\end{equation}
\end{widetext}
where the $a_m^{(p)}, b_m^{(p)}$ are, respectively, the coefficients of the series expansion of $\frac{1}{(1+r)}$  and
$\frac{1}{{(1+r)}^4}$ over the interval $I_p=[r_p, r_{p+1}]$. \\
The results are displayed in the table below proving once again the remarkable 
versatility of the CFM method.

\begin{table}[!ht]
\begin{center}
\begin{tabular}{ c c c c }
\hline  CFM &  Index & Exact & Ratio \\ \hline
  0.004 440 783 & 1 & 0.004 444 444&  0.9992   \\
  0.017 917 823 & 2 & 0.017 777 776&  1.0078   \\
   0.040 324 728 &  3 & 0.039 999 999&  1.0081   \\
   0.071 671 344 & 4 & 0.071 111 105&  1.0078   \\
  0.111 964 293 & 5 & 0.111 111 104&  1.0076   \\
   0.161 208 108 &  6 &  0.159 999 996&  1.0075   \\
   0.219 427 973 &  7 &  0.217 777 774&  1.0075   \\
   0.286 566 615 &  8 &  0.284 444 422&  1.0074   \\
  0.362 612 218 &  9 & 0.359 999 985&  1.0072   \\
   0.447 584 212 & 10 & 0.444 444 418&  1.0070   \\
  0.541 484 714 & 11 &  0.537 777 781&  1.0068   \\
   0.644 318 163 & 12 & 0.639 999 986&  1.0067   \\
   0.756 076 396 & 13 & 0.751 111 090&  1.0066   \\
   0.876 762 569 & 14 & 0.871 111 095&  1.0064   \\
   1.006 386 52 & 15 &  0.999 999 94&   1.0063   \\
\hline
\end{tabular}
\caption{Quantum levels of the infinite square well potential given by the CFM along with exact
results and the ratio of the CFM to the exact eigenvalue. The well width $w$ is chosen in a way such that
the eigenvalue is 1 when the number of levels is 15.} 
\label{well}
\end{center}
\end{table}

After proving the success of the CFM over a finite interval we move on
to the standard application over the usual infinite interval $[0,\infty[$. \\

Double-minimum Potential Well (DPW) problems are interesting to solve as they arise
in many  areas of Atomic, Molecular and even Solid State physics.
They are encountered in highly excited Rydberg states of atoms in crossed electric and
magnetic fields and in certain molecular potential curves.
When two-dimensional electron layers (such as in semiconducting heterostructures) 
are placed in perpendicular electric and magnetic fields, a potential well with
two minima, for the electronic motion normal to the surface, arises. \\

A DPW can be symmetric or asymmetric and one has to adapt in each case
the appropriate boundary condition imposed by the CFM (see Appendix 
on matching and boundary conditions).

An interesting DPW is the symmetric Double Gaussian potential
investigated by Hamilton and Light \cite{Ham}. It is given by:
$$
    V(r)=-D [ \exp(-\Omega {(r-r_{a})}^2)+\exp(-\Omega {(r+r_{a})}^2)]
$$

The values of the parameters: $D, \Omega, r_{a}$ are respectively: 12.0,0.1,5.0 in standard a.u
such that $\hbar=1, \mu=m_e=1$.

The results shown for all the 24 levels in table~\ref{dgw} prove unambigously the accuracy of the CFM method
and its high level of competitivity with respect to the somehow sophisticated method used by  
Hamilton and Light \cite{Ham} based on Distributed Gaussian Basis sets inspired from Quantum
 Chemistry Techniques.

\begin{table}[!ht]
\begin{center}
\begin{tabular}{ l c l }
\hline
  CFM & Index &  Hamilton and Light \\
\hline
     -11.250 418 469    &	0   &  -11.245 199 313   \\	
     -9.779 202 594    & 2   &  -9.773 496 902	 \\	
     -8.387 701 732    & 4   & -8.381 307 510 \\		
     -7.079 415 041   &	6   & -7.072 039 562 \\		
     -5.858 805 474   &	8   & -5.849 958 02  \\	
     -4.732 231 858   &	10  & -4.720 829 36 \\	
     -3.709 907 559    & 12  & -3.694 518 38 \\
     -2.807 436 691   &	14  & -2.798 251 92  \\	
     -2.022 064 904   &	16  & -2.089 661 3 \\		
     -1.293 090 067    & 18  & -1.462 202 9 \\	
     -0.601 483 056   & 20  & -0.771 081  \\	
     -0.067 153 689    & 22  & -0.177 181 \\
\hline
\hline
      -11.250 421 409   &	 1  & -11.245 199 313 \\
      -9.779 225 834    &	 3  &  -9.773 496 902  \\
     -8.387 719 137    &	 5  &  -8.381 307 491  \\
     -7.079 412 929     &	 7  & -7.072 038 846  \\
     -5.858 811 221    &	 9  & -5.849 940 0 \\
     -4.732 171 001   &	 11 & -4.720 509 6  \\
     -3.709 113 861  &	 13 & -3.690 475 6  \\
     -2.801 628 760   &	 15 & -2.763 219 7  \\
     -2.000 566 637   &	 17 & -1.924 577 \\
     -1.255 332 005   &	 19 & -1.149 254 \\
     -0.561 216 170   &	 21 & -0.457 88 \\
     -0.045 810 537  &	 23 & -0.003 41 \\     
      \hline
\end{tabular}
\end{center}
\caption{Computed eigenvalues for the symmetric double Gaussian well potential. The even index results
are shown separately from the odd index since they correspond to different symmetries. 
The numbers at left are the levels computed 
with the CFM; on the right the levels obtained by Hamilton and Light \cite{Ham}. Note
the deterioration of accuracy of the Hamilton and Light results as the index increases 
because of the approach of the continuum.}
\label{dgw}
\end{table}

Johnson has studied an asymmetric DWP consisting of the sum of a Morse
and a Gaussian resulting in the expression:

$$
V(r)=D{[1-\exp(-B(r-r_{a}))]}^2+A \exp{(-C(r-r_{b})}^2) 
$$       

The values of the parameters $A,B,C,D,r_a,r_b$ are (following Johnson~\cite{Johnson})
 in (cm$^{-1}$, \AA \hspace{1mm} system of units) are: 10$^4$ cm$^{-1}$,    1.54 \AA$^{-1}$, 
 200.0 \AA$^{-2}$,    31250.0 cm$^{-1}$,   1.5  \AA,    1.6  \AA  \hspace{1mm} respectively.

Eigenvalues for the asymmetric double minimum 
potential problem are given in table~\ref{DWP2} and  a
comparison between Johnson's~\cite{Johnson} results and the CFM are displayed below.

\begin{table}[!ht]
\begin{center}
\begin{tabular}{c c c}
\hline Index& Johnson & CFM \\
\hline
0 & 1302.500  &         1302.498 972    \\
1 & 3205.307  &  	3205.303 782   \\ 
2 & 4227.339  &  	4227.336 543    \\
3 & 5144.251  &  	5144.243 754  \\
4 & 6064.241  &  	6064.225 881   \\ 
5 & 7092.679  &  	7092.664 815  \\
6 & 7614.622  &  	7614.603 506  \\
7 & 8911.545  &  	8911.513 342    \\ 
8 & 9095.696  &  	9095.679 497   \\  
9 & 10208.350  &    10208.318 142    \\ 
10 & 10869.289  &    10869.255 077    \\ 
11 & 11482.475  &    11482.457 956    \\ 
12 & 12353.799  &    12353.766 422    \\  
13 & 12972.473  &    12972.453 117    \\ 
14 & 13690.455  &    13690.436 602   \\  
15 & 14435.350  &    14435.321 044   \\ 
\hline
\end{tabular}
\caption{Eigenvalues in cm$^{-1}$ of the Johnson asymmetric DWP consisting of the
sum of a Morse and a Gaussian potentials $V(r)=D{[1-\exp(-B(r-r_{a}))]}^2+A \exp{(-C(r-r_{b})}^2)$.
with $A= 10^4 \rm{cm}^{-1},   B= 1.54 \AA^{-1}, 
C= 200.0 \AA^{-2},   D= 31250.0 \rm{cm}^{-1}, r_a =  1.5  \AA,  r_b=  1.6  \AA $.
Johnson~\cite{Johnson} results are compared to the CFM.}
\label{DWP2}
\end{center}
\end{table}

\section{Conclusions}

In this review, we have shown that the CFM is a very powerful and
accurate method that is able to solve a large variety of quantum problems such as:

\begin{itemize}
\item Energy levels for regular and singular potentials (radial case)
\item Pseudo-potential estimation (parametric potential) from spectroscopic data
\item Accurate Phase shift evaluation for regular and singular potentials
\item Test of the Bohr Correspondence Principle
\item Vibrational energy levels of Cold Molecules and application to the $^{23}{\rm Na}_2$
molecule in the $0^-_g$ and $1_u$ electronic states. The Lennard-Jones molecule case is also
studied.
\item Local and  Non-Local Exchange problems.
\item Energy levels for regular and singular potentials (general 1D case)
\end{itemize}

Its mathematics is quite subtle since it enables one to find the Vibrational
spectra of tenuous molecules where energies and distances are so remote from
the ordinary short-range molecules case that special techniques 
should be developed in order to avoid numerical instabilities and uncertainties.\\

The CFM has been tested succesfully in Long-range and Short-range potentials 
for Atomic and Molecular states and gives accurate
results for bound and free states ~\cite {Kobeissi82, Kobeissi90, Kobeissi91}. 
The tunable accuracy of our method allows to evaluate eigenvalues
close to the ground state as well as close to highly excited states near
the continuum limit to a large number of digits without any extrapolation.\\

The CFM compares favorably with many different sophisticated techniques 
such as those developed, for instance (to cite a few), by 
Raptis {\it et al.}~\cite{Raptis85, Raptis87}, Johnson ~\cite{Johnson},
Hamilton  {\it et al.}~\cite{Ham} and Rawitscher {\it et al.}~\cite{Raw}. The CFM
approach remains the same despite the wide variability of the mentioned problems.  

The VSCA integration method used gives the right number of all the levels
and the variation of the eigenvalue function $F(E)$ definitely determines the
total number of levels. Generally it requires performing analytically Taylor
series expansion to any order of an arbitrary potential function that might 
require the combination of numerical, symbolic manipulation, functional
fitting and analytic function continuation techniques 
(see refs~\cite{Fornberg1,Fornberg2}). Despite these challenges, 
the results of the VSCA are rewarding.\\

The use of the RK4 and the VSCA methods jointly paves the way to a precise
comparative evaluation of the accuracy of the spectra obtained. In practice,
the RK4 method can be used in optimisation problems whereas the VSCA is more
adapted to the direct evaluation of the spectra.\\

Since the CFM bypasses the calculation of the
eigenfunctions, it avoids losing accuracy associated with the
numerical calculation specially with rapidly oscillating wave functions
of highly excited states. This is specially needed in the study of the 
sensitive problem of determining the vibrational spectra of cold molecules.

We did not consider molecular rotation, nevertheless the CFM
is adapted to solve accrately Ro-Vibrational problems as well as any RSE 
diagonalisation problem (see for instance ref.~\cite{Kobeissi90}).\\ 

Regarding spin dependent scattering problems we applied the CFM to local 
and non-local exchange integrating exactly the coupled equations originating from
integro-differential equations.
We benchmarked and tested our integration method at extremely small
energies for momentum $k \ge 10^{-8}$ and found that it performs again very well 
against stability error $\epsilon_S$ and truncation error $\epsilon_T$.
Our results were validated with a highly 
accurate method, the recently developed S-IEM \cite{Raw}. The results we find  
(phaseshifts and scattering lengths) agree with the  S-IEM results for 
the singlet, triplet cases for an arbitrarily small value of momentum $k$ ( $k \ge 10^{-8}$). \\ 

In conclusion, the (single channel and multichannel) CFM enables one to tackle precisely weakly
bound states in atoms and (simple and cold) molecules as well as 
low-energy scattering of atoms 
and molecules and can be extended straightforwardly to Bose-Einstein
 condensation problems \cite {Trost98}.\\

{\bf Acknowledgements}: This work is dedicated to the memory of Hafez Kobeissi 
(1936-1998), our past teacher (CT and KF) and colleague 
who laid the foundations of the CFM. Part of this work was performed on the IDRIS
(Cray and IBM Clusters) machines at the CNRS (Orsay).\\
We would like to acknowledge helpful correspondance with Jeff Cash (Imperial College),
Ronald Friedman (Purdue), Bengt Fornberg (Caltech) and John W. Wilkins (Ohio state).\\

\begin{center}
{\bf APPENDIX}
\end{center}

{\bf Atomic and other units} \\

In atomic and molecular physics, it is convenient to use the elementary
charge $e$, as the unit of charge, and the electron mass $m_e$
as the unit of mass (although for some purposes
the proton mass, $m_p$ , or the unified mass unit amu,
is more convenient). Electrostatic forces and energies in atoms are
proportional to $e^2/4\pi \epsilon_0$ , which has dimensions
$ML^3T^{-2}$, and another quantity that appears all over in
quantum physics is $\hbar$ which has dimensions $ML^2T^{-1}$
; so it is convenient to choose units of length and time such that
 $4\pi \epsilon_0=1$ and $\hbar=1$.

The atomic unit of length is then (by dimensional analysis alone)

\begin{equation}
a_B=\frac{e^2}{m_e (e^2/4\pi \epsilon_0)}
\end{equation}

This is called the Bohr radius, or simply the bohr (0.529 \AA), 
because in the "Bohr model" the radius of the smallest orbit for an electron circling a
fixed proton is $(1+m_e/m_p)a_B$.  In the full quantum theory
the particles do not follow an orbit, but the expectation value of the
electron-proton distance in the Hydrogen ground state is exactly
$(1+m_e/m_p)a_B$).

The atomic unit of energy is the Hartree (27.2 eV) given by:
\begin{equation}
E_h= \frac{e^2}{4\pi \epsilon_0}\frac{1}{a_B}={(\frac{e^2}{4\pi \epsilon_0})}^2\frac{m_e}{\hbar^2}
\end{equation}

The unit of time is $\hbar/E_h$.

The Hartree is twice the ground state energy of the Hydrogen atom
$\frac{1}{2}{(1+m_e/m_p)}^{-1}E_h$ equal to the Rydberg (13.6 eV). 
In atomic and molecular spectroscopy, one uses rather the cm$^{-1}$ an energy
corresponding to a wavelength of 1cm or sometimes a frequency unit, the Hz.
We refer the reader to the table below giving the conversion factors between the
different energies.

\begin{widetext}

\begin{table}[!h]
\begin{center}
\begin{tabular}{  l  r  c c  c  c   }
\hline
       &    & J & eV & Hz & cm$^{-1}$  \\
\hline
1 J  &= & 1 &  6.24151.10$^{18}$ &  1.50919.10$^{33}$ & 5.03411.10$^{22}$  \\
\hline
1 eV & = & 1.60219.10$^{-19}$ & 1 & 2.41797.10$^{14}$ & 8.06547.10$^{3}$ \\
\hline
1 Hz & = & 6.62619.10$^{-34}$ & 4.13570.10$^{-15}$ & 1 & 3.33564.10$^{-11}$  \\
\hline
1 cm$^{-1}$ & = & 1.96648.10$^{-23}$ & 1.23935.10 $^{-4}$ & 2.99792.10$^{10}$ & 1  \\
\hline
\end{tabular}
\end{center}
\end{table}

\end{widetext}

{\bf Matching and boundary conditions} \\

The CFM is based on the extraction of the eigenvalues from the zeroes of the
eigenvalue function $F(E)$ defined from the saturation of the left ($r \rightarrow 0$)
and right ($r \rightarrow +\infty$) functions $l_{-}(E)$ and $l_{+}(E)$ given
by the ratios of the canonical functions $\alpha(r)$ and $\beta(r)$.
This was described previously in the case the boundary conditions are
$y(0)=y(\infty)=0$. 
In the general case we write:
\begin{eqnarray}
y(r)&= y(r_0) \alpha(E;r)  + y'(r_0)\beta(E;r)  \nonumber \\
y'(r)&= y(r_0) \alpha'(E;r)  + y'(r_0)\beta'(E;r)
\label{cano}
\end{eqnarray}

The canonical functions satisfy the conditions:
\begin{equation}
\alpha(E;r_0)=1,\alpha'(E;r_0)=0, \beta(E;r_0)=0,\beta'(E;r_0)=1 
\end{equation}

Let us rewrite the system \ref{cano} at the two boundaries $r=0$ :
\begin{eqnarray}
y(0)&= y(r_0) \alpha(E;0)  + y'(r_0)\beta(E;0)  \nonumber \\
y'(0)&= y(r_0) \alpha'(E;0)  + y'(r_0)\beta'(E;0)
\label{zero}
\end{eqnarray}

and at $r=\infty$:

\begin{eqnarray}
y(\infty)&= y(r_0) \alpha(E;\infty)  + y'(r_0)\beta(E;\infty)  \nonumber \\
y'(\infty)&= y(r_0) \alpha'(E;\infty)  + y'(r_0)\beta'(E;\infty)
\label{infty}
\end{eqnarray}

Extracting from above the left and right ratios:
\begin{eqnarray}
\left.\frac{y'(r_0)}{y(r_0)}\right]_{-} &=\frac{\alpha(E;0)y'(0)-\alpha'(E;0)y(0)}{\beta'(E;0)y(0)-\beta(E;0)y'(0)}   \nonumber \\
\left. \frac{y'(r_0)}{y(r_0)}\right]_{+} &=
\frac{\alpha(E;\infty)y'(\infty)-\alpha'(E;\infty)y(\infty)}{\beta'(E;\infty)y(\infty)-\beta(E;\infty)y'(\infty)} 
\end{eqnarray}

we can tackle several types of boundary conditions providing below the energy functions $l_{-}(E)$ and $l_{+}(E)$
that determine the eigenvalue function: 
\begin{equation}
F(E)=l_{+}(E)-l_{-}(E)= \left[ \frac{y'(r_0)}{y(r_0)}\right]_{+} - \left[\frac{y'(r_0)}{y(r_0)}\right]_{-}
\end{equation}

Some of the boundary conditions are, for instance:

\begin{enumerate}
\item Standard (Dirichlet) case:  \\
The boundary conditions $y(0)=y(\infty)=0$ yield:

\begin{eqnarray}
l_{-}(E) = \lim_{r \rightarrow 0} -\frac{\alpha(E;r)}{\beta(E;r)}   \nonumber \\
l_{+}(E) = \lim_{r \rightarrow +\infty} -\frac{\alpha(E;r)}{\beta(E;r)}  
\end{eqnarray}

\item Derivative asymmetric case type I:  \\
The boundary conditions $y(0)=y'(\infty)=0$ yield:

\begin{eqnarray}
l_{-}(E)= \lim_{r \rightarrow 0} -\frac{\alpha(E;r)}{\beta(E;r)}   \nonumber \\
l_{+}(E)= \lim_{r \rightarrow +\infty} -\frac{\alpha'(E;r)}{\beta'(E;r)}  
\end{eqnarray}

\item Derivative asymmetric case type II:  \\
The boundary conditions $y'(0)=y(\infty)=0$ yield:

\begin{eqnarray}
l_{-}(E)= \lim_{r \rightarrow 0} -\frac{\alpha'(E;r)}{\beta'(E;r)}   \nonumber \\
l_{+}(E)= \lim_{r \rightarrow +\infty} -\frac{\alpha(E;r)}{\beta(E;r)}   
\end{eqnarray}

\item Derivative (Neumann type) boundary conditions:  \\
The boundary conditions $y'(0)=y'(\infty)=0$ yield:

\begin{eqnarray}
l_{-}(E)= \lim_{r \rightarrow 0} -\frac{\alpha'(E;r)}{\beta'(E;r)}   \nonumber \\
l_{+}(E)= \lim_{r \rightarrow +\infty} -\frac{\alpha'(E;r)}{\beta'(E;r)}   
\end{eqnarray}

\end{enumerate}

These formulas can also be generalised to arbitrary (Cauchy or mixed type) boundary conditions:
\begin{eqnarray}
a_1 y(0) + b_1 y(0)&=c_1 \nonumber \\
a_2 y(\infty) + b_2 y(\infty)&=c_2
\end{eqnarray}

and to the multichannel case. For instance, in the 1D symmetric Double Gaussian potential
the boundary conditions correspond to above case No.3 whereas the asymmetric case corresponds to
case No. 2.


\begin{thebibliography}{9}

\bibitem{Aymar96} M.~Aymar, C.~H. Green, and E.~Luc-Koenig, Rev. Mod. Phys. {\bf 68}, 1015 (1996).
\bibitem{Aymar}M. Aymar and M. Crance: J. Phys. {\bf B 13}, p. 2527-2544 (1980).
\bibitem{bar} Bar Shalom A 1983 \textit{PhD. Thesis} University of Jerusalem.
\bibitem{Bayliss} Bayliss W E and Peel S J 1982 \textit{Comput. Phys. Commun}. \textbf{25} 7
\bibitem{Boisseau00} C. Boisseau, E. Audouard, J. Vigu\'e and V. V. Flambaum: Eur. J. Phys. {\bf D12}, 199 (2000).
\bibitem{Boisseau98} C. Boisseau, E. Audouard, J. Vigu\'e: Europhys. Lett.  {\bf 41}, 349 (1998).
\bibitem{Bransden}  Bransden B H and Noble C J 1976 \textit{J. Phys. B: At. Mol. Opt. Phys.} \textbf{9} 1507.
\bibitem{Bray} Bray I 1994 \textit{Phys. Rev. Lett.} \textbf{73} 1088.
\bibitem{Broyden} Broyden, CG: Mathematics of Computation, {\bf 19}, 557 (1965). 
\bibitem{Burgess} Burgess A 1963 \textit{Proc. Phys. Soc.} \textbf{81} 442.
\bibitem{Crubellier99} A. Crubellier, O. Dulieu, F. Masnou-Seeuws, M. Elbs, H. Knockel and E. Tiemann, Eur. Phys. J. {\bf D6}, 211 (1999).
\bibitem{Drachman} Drachman R J and Temkin A 1972 \textit{Case Studies in Atomic Physics} \textbf{II}, ed. E W McDaniel and M R C McDowell, North Holland, 399
\bibitem{Fakhreddine06}  Fakhreddine K, Tweed R., Nguyen G., Tannous C., Langlois J. and Robaux O., \textit{Can. J. Phys.} June 2006
\bibitem{Fakhreddine94} Fakhreddine K and Kobeissi H 1994 \textit{Int. J. Quantum Chem.} \textbf{49} 773.
\bibitem{Fakhreddine99}  Fakhreddine K, Kobeissi H and Korec M 1999 \textit{Int. J. Quantum Chem.} \textbf{73} 325.
\bibitem{Fornberg1} B. Fornberg, ACM Trans. on Mathematical Software Vol. \textbf{7}, No 4, 512 (1981).
\bibitem{Fornberg2} B. Fornberg, ACM Trans. on Mathematical Software Vol. \textbf{7}, No 4, 542 (1981).
\bibitem{Friedman} R. S. Friedman and M. J. Jamieson: CPC \textbf{62}, 53 (1991).
\bibitem{Furness} Furness J B and McCarthy I E 1973 \textit{J. Phys. B: At. Mol. Opt. Phys.} \textbf{6} 2280.
\bibitem{Gao99} Bo Gao Phys. Rev. Lett. 83, 4225 (1999).
\bibitem{Green69} A.~E.~S. Green, D.~L. Sellin and  A.~S.Zachor, Phys. Rev. {\bf 184}, 1 (1969).
\bibitem{Ham} I P Hamilton and J C Light: J. Chem. Phys. \textbf{84}, 306 (1986).
\bibitem{Henry} Henry R.J. W., Rountree S. P. and Smith E.R. 1981 \textit{Comp. Phys. Comm.} \textbf{23} 233.
\bibitem{Hibbert82} H.~Hibbert, Adv. At. Mol. Phys. {\bf 18} 309 (1982).
\bibitem{Johnson} B. R. Johnson: J. Chem. Phys. \textbf{67}, 4086 (1977).
\bibitem{Jones96} K. M. Jones, P. S. Julienne, P. D. Lett, W. D. Philips, E. Tiesinga and C. J. Williams, Eur. Phys. Lett. {\bf 35}, 85 (1996).
\bibitem{Jungen} C. Jungen, Molecular Applications of Quantum Defect Theory, Institute of Physics Publishing (1996).
\bibitem{Klapisch71}  M.~Klapisch, Comp. Phys. Comm. {\bf 2},  239 (1971).
\bibitem{Kobeissi82} H. Kobeissi, J. Phys. B {\bf 15}, 693 (1982).
\bibitem{Kobeissi88} H. Kobeissi and M. Kobeissi, J. Comp. Phys. {\bf 77}, 501 (1988).
\bibitem{Kobeissi90} H. Kobeissi, K. Fakhreddine and M. Kobeissi, Int. J. Quantum Chemistry, {\bf XL}, 11 (1990).
\bibitem{Kobeissi91a} Kobeissi H and Fakhreddine K 1991a \textit{J. Phys. II (France)} \textbf{1} 899.
\bibitem{Kobeissi91b} Kobeissi H and Fakhreddine K 1991b \textit{J. Comput. Phys.} \textbf{95} 505.
\bibitem{Kobeissi91c} Kobeissi H Fakhreddine K and Kobeissi M 1991 \textit{Int. J. Quantum Chem.}\textbf{XL} 11.
\bibitem{Kobeissi91} H. Kobeissi, K. Fakhreddine  J. Physique II (France) {\bf 1}, 38 (1991).
\bibitem{LeRoy70} R. J. LeRoy and R. B. Bernstein, J. Chem. Phys. {\bf 52}, 3869 (1970).
\bibitem{Maceachran} R. P. McEachran, A. D. Stauffer, J. Phys. {\bf B 16}, 4023 (1983). 
\bibitem{Mccarthy} I. E. McCarthy, Aust. J. Phys. {\bf 48}, 1 (1995). 
\bibitem{McDowella} McDowell M R C, Morgan L and Myerscough V P 1974a \textit{Comput. Phys. Commun.} \textbf{7} 38.
\bibitem{McDowellb} McDowell M R C, Myerscough V P and Morgan L 1974b \textit{J. Phys. B: At. Mol.Opt. Phys.} \textbf{24} 657.
\bibitem{Moore} C. E. Moore, Atomic energy levels, NBS Publications (1971). 
\bibitem{Movre77} M. Movre and  G. Pichler, J. phys. {\bf B10}, 2631 (1977).
\bibitem{Numerov} Numerov B 1933 \textit{Obser. Cent. Astrophys. (Russ.)} \textbf{2} 188.
\bibitem{Pan} C. Pan and A. F. Starace, Phys Rev {\bf A 45}, 4588 (1992). 
\bibitem{Peterkop} R. K. Peterkop, \textit{ Theory of Ionization of Atoms By Electron Impact} (Colorado Associated University Press, Boulder) (1977). 
\bibitem{Raptis85} A. D. Raptis and J.R. Cash:  CPC  {\bf 36}, p. 113-119 (1985).
\bibitem{Raptis87} A. D. Raptis and J.R. Cash: CPC {\bf 44}, p. 95-103  (1987).
\bibitem{Rau}A. R. P. Rau and M. Inokuti:  Am. J. Physics, {\bf 65}, p. 221-225 (1997).
\bibitem{Raw} Rawitscher G.H, Kang S-Y and Koltracht I. 2003 \textit{J. Chem. Phys.} \textbf{118} 9149.
\bibitem{Recipes} Numerical Recipes in C: The Art of Scientific Computing, W. H. Press, W. T. Vetterling, S. A. Teukolsky and B. P. Flannery, Second Edition, page 389, Cambridge University Press (New-York, 1992).
\bibitem{Riley} M.E. Riley and D.G. Trular,J. Chem. Phys. {\bf 63}, 2182 (1975).
\bibitem{Roche1}  P.J.P. Roche, S. Kawano, C.T. Whelan, J. Rasch, H. R. J. Walters, R. J. Allan, J. Langlois and C. Tannous 2001, Chapter 7, pp. 81-90, in
\textit{Many-Particle Spectroscopy of Atoms, Molecules, Clusters and Surfaces} edited by J. Berakdar and J. Kirschner, Kluwer Academic/Plenum Publishers (New-York, 2001). 
\bibitem{Roche} C. Jungen, A.L. Roche, M Arif, Phil. Trans. {\bf A 355}, 2520 (1997).
\bibitem{Rouet} Rouet F, Tweed R J and Langlois J J. Phys. B: At. Mol. Opt. Phys. \textbf{29} 1767  (1996).
\bibitem{Rouvellou} B. Rouvellou, S. Rioual, J. Roeder, A. Pochat, J. Rasch, C. T. Whelan, H. R. J. Walters and R. J. Allan, Phys. Rev. {\bf A 57}, 3621 (1998).
\bibitem{Scott} Scott T and McDowell M R C 1975 \textit{J. Phys. B: At. Mol. Opt. Phys.}  \textbf{8} 1851.
\bibitem{Stwalley78} W.C. Stwalley, Y.H. Uang and G. Pichler, Phys. Rev. Lett. {\bf 41}, 1165 (1978).
\bibitem{Szydlik74} P.~P. Szydlik and A.~E.~S.Green, Phys. Rev. {\bf A9}, 1885 (1974).
\bibitem{Tannous99} C. Tannous, K. Fakhreddine and J. Langlois, J. Phys. IV France {\bf 9} Pr6-71 (1999).
\bibitem{Trost98} J. Trost, C. Eltschka and H. Friedrich: J. Phys. {\bf B31}, 361 (1998).
\bibitem{Weigold} I. E. McCarthy and E. Weigold, \textit{Electron-atom collisions}, Cambridge University Press (1995). 
\bibitem{Whelan} Whelan C. T. 1999 in \textit{New Directions in Atomic Physics}, edited by C.T. Whelan, R. M. Dreizler, J. H. Macek and H. R. J. Walters (Kluwer/Plenum, New York).
\bibitem{Winkler} K. D. Winkler, D. H. Madison and H. P. Saha J. Phys. {\bf B 32}, 1987 (1999).
\end{thebibliography}
\end{document}